\documentclass[12pt]{article}
\usepackage{amsfonts,graphicx,amsmath,amsthm,amssymb,epsfig}
\usepackage{float}
\allowdisplaybreaks
\begin{document}

\title{\bf Charged Anisotropic Tolman IV Solution in Matter-Geometry Coupled Theory}
\author{M. Sharif$^1$ \thanks{msharif.math@pu.edu.pk} and Tayyab Naseer$^{1,2}$ \thanks{tayyabnaseer48@yahoo.com}\\
$^1$ Department of Mathematics and Statistics, The University of Lahore,\\
1-KM Defence Road Lahore, Pakistan.\\
$^2$ Department of Mathematics, University of the Punjab,\\
Quaid-i-Azam Campus, Lahore-54590, Pakistan.}

\date{}
\maketitle

\begin{abstract}
This paper discusses the interior distribution of several
anisotropic star models coupled with an electromagnetic field in the
context of $f(\mathcal{R},\mathcal{T},\mathcal{Q})$ gravity, where
$\mathcal{Q}=\mathcal{R}_{\beta\xi}\mathcal{T}^{\beta\xi}$. In this
regard, a standard model of this modified gravity is taken as
$\mathcal{R}+\nu_{3}\mathcal{R}_{\beta\xi}\mathcal{T}^{\beta\xi}$,
where $\nu_{3}$ symbolizes an arbitrary coupling constant. We assume
a charged spherically symmetric metric that represents the interior
geometry of compact quark stars and develop the corresponding
modified field equations. These equations are then solved with the
help of metric potentials of Tolman IV spacetime and a linear bag
model equation of state. We consider the experimental data (i.e.,
radii and masses) of different quark models such as SMC X-4,~SAX J
1808.4-3658,~Her X-I and 4U 1820-30 to analyze how the charge and
modified corrections affect their physical characteristics. The
viability and stability of the resulting model is also checked for
the considered star candidates with two different values of
$\nu_{3}$. We conclude that only two models, Her X-I and 4U 1820-30
show stable behavior in this modified framework for both values of
the coupling constant.
\end{abstract}

\section{Introduction}

A widely accepted theory of gravitation is considered as general
relativity ($\mathbb{GR}$) that addresses various challenges related
to cosmic evolution. However, it is still not sufficiently enough to
explain an accelerated expansion of the universe properly. In view
of this, various modifications to $\mathbb{GR}$ have been proposed
that were claimed to handle mystifying issues (i.e., the dark matter
and expeditious expanding cosmos etc.) in some better way. It was
pointed out by several astrophysical experiments that there exists a
heavy amount of a mysterious force having large negative pressure,
and thus causes such accelerated expansion, called dark energy. The
first ever extension to $\mathbb{GR}$ has been proposed by modifying
the geometric part of the Einstein-Hilbert action in which the Ricci
scalar $\mathcal{R}$ was swapped by its generic function, thus
termed the $f(\mathcal{R})$ gravity \cite{1}. Multiple models of
this modified theory have been discussed in the literature and it
was deduced that they produce viable and stable self-gravitating
models \cite{2}-\cite{2f}.

An interesting notion in the framework of $f(\mathcal{R})$ gravity
was initially introduced by Bertolami \emph{et al.} \cite{10}, in
which they coupled the effects of matter and geometric terms in the
matter Lagrangian. This idea prompted many scientists to study how
such a coupling affects the interior of a self-gravitating object.
The interesting results were found after an extensive study that
urged researchers to generalize the concept of coupling at the
action level. This has been done by Harko \emph{et al.} \cite{20}
for the first time, who introduced $f(\mathcal{R},\mathcal{T})$
gravitational theory, in which $\mathcal{T}$ indicates trace of the
energy-momentum tensor $(\mathbb{EMT})$. However, such a generalized
coupling immediately makes this modified theory non-conserved,
opposing $f(\mathcal{R})$ gravity as well as $\mathbb{GR}$. Several
impressive astrophysical outcomes have been observed while studying
this extended theory \cite{21}-\cite{21f}.

The $f(\mathcal{R},\mathcal{T})$ gravity was further extended by
Haghani \emph{et al.} \cite{22} who added a term
$\mathcal{Q}\equiv\mathcal{R}_{\beta\xi}\mathcal{T}^{\beta\xi}$ in
the preceding functional. The viability of three distinct models
have been studied thoroughly in this framework. The main reason to
suggest this theory is that the $f(\mathcal{R},\mathcal{T})$ gravity
fails to entail a non-minimal interaction on test particles when a
traceless fluid distribution is assumed. However, this theory does
entail even in this scenario due to the term
$\mathcal{R}_{\beta\xi}\mathcal{T}^{\beta\xi}$. The galactic
rotation and exponential expansion era of our universe can also be
studied with the help of this theory. Sharif and Zubair \cite{22a}
chosen density as well as pressure-dependent matter Lagrangian, and
discussed the laws of black hole thermodynamics for the models such
as $\mathcal{R}+\nu_3\mathcal{R}_{\beta\xi}\mathcal{T}^{\beta\xi}$
and
$\mathcal{R}(1+\nu_3\mathcal{R}_{\beta\xi}\mathcal{T}^{\beta\xi})$.
The physical viability of these two models has also been checked and
it was deduced that negative coupling constant does not satisfy the
weak energy bounds \cite{22b}.

The modified cosmological solutions have been formulated by Odintsov
and S\'{a}ez-G\'{o}mez \cite{23}, and they also verified that the
$\Lambda$CDM model is supported by this extended gravity. Baffou
\emph{et al.} \cite{25} explored the stable regions of the solutions
(obtained by solving Friedmann equations numerically) with respect
to two different $f(\mathcal{R},\mathcal{T},\mathcal{Q})$ models.
Sharif and Waseem \cite{25a,25b} discussed the isotropic/anisotropic
structures by taking the matter Lagrangian to be the negative
pressure in this theory and obtained unaccepted physical results.
Yousaf and his collaborators \cite{26}-\cite{26e} split the Riemann
tensor (engaging modified $\mathbb{EMT}$) to obtain some scalars
which are used to study the interior configuration of
self-gravitating spherical systems. The simplest possible
evolutionary modes along with the complexity factor for non-static
cylindrical systems have also been studied \cite{27aa,27aaa}.
Recently, we have used the decoupling approach and a quark model
equation of state to obtain feasible solutions of modified field
equations \cite{27}-\cite{27a2}.

Researchers have done extensive analysis in the scenario of
$\mathbb{GR}$ and modified theories of gravity to check whether the
presence of charge in the interior of a compact star makes it
physically more relevant or not. An exterior geometry represented by
the Reissner-Nordstr\"{o}m metric has been considered by Das
\emph{et al.} \cite{27a}, with the help of which they calculated the
solution of a charged model at the spherical interface. Sunzu
\emph{et al.} \cite{27b} employed a well-known relation between
radius and mass of a strange quark star whose interior was
influenced from an electromagnetic field, and discussed its physical
properties. An extensive analysis has been done on the physical
relevancy of charged compact self-gravitating objects, and it was
deduced that the presence of charge makes these systems more stable
\cite{27e}-\cite{27i}.

Neutron stars have such intense gravity that they crush electrons
and protons together into neutrons. The whole star is made of
neutrons, inside and out. If more mass is added to the neutron star,
it will be impossible to hold even the neutrons together, and the
whole object collapses into a black hole. An in-between stage of
neutron stars and black holes is the quark star that has too much
mass in its center for the neutrons to hold their atomness, however,
not enough to completely collapse into a black hole. Quarks are
considered as elementary particles that come in six different types
such as up, down, top, bottom, charm and strange. The two flavors
namely `up' and `down' quarks are squeezed together into `strange'
quarks. Since it is composed of `strange' quarks, physicists name
this `strange matter'. The conjecture that strange matter may be the
absolute ground state of strongly interacting matter resulted in the
possible existence of strange stars \cite{1adf,1adg}. It has been
shown in recent studies that the compact objects associated with the
x-ray pulsar such as Her X-I and 4U 1820-30 are good candidates for
strange stars \cite{1adh,1adi}. Moreover, the charge has been
estimated inside several quark candidates and found to be within the
interval $(1.5453\times10^{18}C,7.6271\times10^{19}C)$ \cite{1adj}.
It was also deduced that minimum and maximum charge corresponds to
the stars Her X-I and SMC X-I, respectively.

Some constraints have been introduced in the literature that
interlink different physical variables representing
isotropic/anisotropic interiors such as pressure and energy density.
Among such constraints, one is the $\mathbf{MIT}$ bag model equation
of state ($\mathbf{E}$o$\mathbf{S}$) that portrays the quark'
interior \cite{27a}. It was observed that only this
$\mathbf{E}$o$\mathbf{S}$ is feasible to calculate the compactness
of strange quark systems like Her X-1, RXJ 185635-3754, 4U 1728-34,
4U 1820-30, PSR 0943+10 and SAX J 1808.4-3658, etc. because an
$\mathbf{E}$o$\mathbf{S}$ for neutron stars flunk to do so
\cite{33a}. The bag constant involving in $\mathbf{MIT}$
$\mathbf{E}$o$\mathbf{S}$ helps to determine the difference between
the true and the false state of a vacuum. An interior fluid
configuration of quark bodies has been extensively studied with the
help of this model \cite{33b}-\cite{34a}. Demorest \emph{et al.}
\cite{34b} considered a strange star PSR J1614-2230 and discussed
its several fundamental characteristics. They found $\mathbf{MIT}$
bag model to be the only suitable candidate to discuss a family of
such heavy objects. Rahaman \emph{et al.} \cite{35} adopted the same
model and explored mass as well as other physical properties of
compact systems through interpolating technique.

Various approaches have been employed to develop solution to the
highly complicated field equations in any theory of gravitation
(either $\mathbb{GR}$ or modified gravity). For instance, a specific
$\mathbf{E}$o$\mathbf{S}$ or developed forms of the metric
functions, etc. can be used in order to do this. One of these
well-known metric potentials is the Tolman IV spacetime that has
been adopted by various researchers to obtain physically acceptable
compact models in different theories of gravity. Murad and Fatema
\cite{36} developed some anisotropic models by using this solution
and showed their physical relevancy through graphical analysis. Bhar
\emph{et al.} \cite{37} investigated physical characteristics
corresponding to various compact structures by using Tolman IV
spacetime and found their resulting solution to be singularity-free
and viable. Malaver \cite{37b} adopted the same solution and
developed a physically feasible charged model to check how an
electromagnetic field affects it. This spacetime has also been
coupled with the complexity factor of anisotropic fluid distribution
through which an acceptable compact model is developed in the
context of $\mathbb{GR}$ as well as Brans-Dicke theory
\cite{37c,38c}. A charged realistic isotropic solution was
formulated and graphically interpreted with respect to some
particular stars such as SAX J1808.4-3658, Her X-1 and 4U 1538-52
\cite{37da}.

In this article, we discuss the interior fluid configuration of the
charged anisotropic quark stars in modified
$f(\mathcal{R},\mathcal{T},\mathcal{Q})$ theory of gravity. The
format of this paper is given as follows. Section \textbf{2} defines
some fundamental entities of this modified gravity and establishes
the field equations for a specific model given by
$\mathcal{R}+\nu_3\mathcal{R}_{\beta\xi}\mathcal{T}^{\beta\xi}$. We
then use the metric components of Tolman IV spacetime and
$\mathbf{MIT}$ bag model $\mathbf{E}$o$\mathbf{S}$ to solve the
field equations. Section \textbf{3} calculates the unknown constants
with the help of the boundary conditions. We further explore the
effects of this theory and charge on matter determinants and
stability of compact models in section \textbf{4}. Finally, section
\textbf{5} summarizes all our results.

\section{Modified Theory}

The action corresponding to $f(\mathcal{R},\mathcal{T},\mathcal{Q})$
gravitational theory (with $\kappa=8\pi$) is given by \cite{23}
\begin{equation}\label{g1}
\mathbb{I}=\int\sqrt{-g}
\left\{\frac{f(\mathcal{R},\mathcal{T},\mathcal{Q})}{16\pi}
+\mathbf{L}_{\mathrm{E}}+\mathbf{L}_{m}\right\}d^{4}x,
\end{equation}
where $\mathbf{L}_{\mathrm{E}}$ and $\mathbf{L}_{m}$ being the
Lagrangian densities of the electromagnetic field and fluid
distribution, respectively. The principle of least-action provides
after implementing on Eq.\eqref{g1} as
\begin{equation}\label{g2}
\mathcal{G}_{\beta\xi}=\mathcal{T}_{\beta\xi}^{(\mathrm{ef})}=8\pi\bigg\{\frac{1}{f_{\mathcal{R}}-\mathbf{L}_{m}f_{\mathcal{Q}}}
\left(\mathcal{T}_{\beta\xi}+\mathrm{E}_{\beta\xi}\right)
+\mathcal{T}_{\beta\xi}^{(\mathrm{cr})}\bigg\},
\end{equation}
where the Einstein tensor $\mathcal{G}_{\beta\xi}$ represents the
spacetime geometry and $\mathcal{T}_{\beta\xi}^{(\mathrm{ef})}$ is
the total $\mathbb{EMT}$ of this extended theory. Also,
$\mathcal{T}_{\beta\xi}$ and $\mathrm{E}_{\beta\xi}$ are the usual
$\mathbb{EMT}$ and the electromagnetic tensor, respectively. The
last term $\mathcal{T}_{\beta\xi}^{(\mathrm{cr})}$ of the above
equation takes the following form
\begin{eqnarray}\nonumber
\mathcal{T}_{\beta\xi}^{(\mathrm{cr})}&=&-\frac{1}{8\pi\bigg(\mathbf{L}_{m}f_{\mathcal{Q}}-f_{\mathcal{R}}\bigg)}
\left[\left(f_{\mathcal{T}}+\frac{1}{2}\mathcal{R}f_{\mathcal{Q}}\right)\mathcal{T}_{\beta\xi}
+\left\{\frac{\mathcal{R}}{2}(\frac{f}{\mathcal{R}}-f_{\mathcal{R}})-\mathbf{L}_{m}f_{\mathcal{T}}\right.\right.\\\nonumber
&-&\left.\frac{1}{2}\nabla_{\eta}\nabla_{\lambda}(f_{\mathcal{Q}}\mathcal{T}^{\eta\lambda})\right\}g_{\beta\xi}
-\frac{1}{2}\Box(f_{\mathcal{Q}}\mathcal{T}_{\beta\xi})-(g_{\beta\xi}\Box-
\nabla_{\beta}\nabla_{\xi})f_{\mathcal{R}}\\\label{g4}
&-&2f_{\mathcal{Q}}\mathcal{R}_{\eta(\beta}\mathcal{T}_{\xi)}^{\eta}
+\nabla_{\eta}\nabla_{(\beta}[\mathcal{T}_{\xi)}^{\eta}f_{\mathcal{Q}}]
+2(f_{\mathcal{Q}}\mathcal{R}^{\eta\lambda}+\left.f_{\mathcal{T}}g^{\eta\lambda})\frac{\partial^2\mathbf{L}_{m}}
{\partial g^{\beta\xi}\partial g^{\eta\lambda}}\right].
\end{eqnarray}
Here, $f=f({\mathcal{R}},{\mathcal{T}},{\mathcal{Q}})$ is partially
differentiated with respect to its arguments and represented by
$f_{\mathcal{R}},~f_{\mathcal{T}}$ and $f_{\mathcal{Q}}$. Also, the
D'Alambert operator is defined as $\Box\equiv
\frac{1}{\sqrt{-g}}\partial_\beta\big(\sqrt{-g}g^{\beta\xi}\partial_{\xi}\big)$
and $\nabla_\beta$ is the covariant derivative.

We recall that the expression of the matter Lagrangian density is
not unique, rather this depends on nature of the matter source of
the universe \cite{22}. Usually, it can be taken in terms of the
energy density, pressure or some scalar field. Since we consider the
charged framework, a more appropriate choice of the matter
Lagrangian in this regard is taken as
$\mathbf{L}_{m}=-\frac{1}{4}\mathcal{Z}_{\beta\xi}\mathcal{Z}^{\beta\xi}$,
giving rise to $\frac{\partial^2\mathbf{L}_{m}} {\partial
g^{\beta\xi}\partial
g^{\eta\lambda}}=-\frac{1}{2}\mathcal{Z}_{\beta\eta}\mathcal{Z}_{\xi\lambda}$
\cite{22}. Further, the Maxwell field tensor can be expressed as
$\mathcal{Z}_{\beta\xi}=\frac{\partial\tau_{\xi}}{\partial
x^{\beta}}-\frac{\partial\tau_{\beta}}{\partial x^{\xi}}$, where the
four potential $\tau_\xi$ is given by
$\tau_{\xi}=\big(\tau(r),0,0,0\big)$. The generic functional of this
gravity results in the violation of the principle of equivalence
owing to the terms representing an arbitrary matter-geometry
coupling. We establish the covariant divergence of $\mathbb{EMT}$
\eqref{g4} which is observed to be non-conserved, (i.e.,
$\nabla_\beta \mathcal{T}^{\beta\xi}\neq 0$). The geodesic motion of
the moving test particles is now altered due to the existence of an
extra force. Mathematically, we have
\begin{align}\nonumber
\nabla^\beta
\mathcal{T}_{\beta\xi}&=\frac{2}{2f_\mathcal{T}+\mathcal{R}f_\mathcal{Q}+16\pi}
\bigg[\nabla_\beta\big(f_\mathcal{Q}\mathcal{R}^{\eta\beta}
\mathcal{T}_{\eta\xi}\big)-\mathcal{G}_{\beta\xi}\nabla^\beta\big(f_\mathcal{Q}\mathbf{L}_m\big)\\\label{g4a}
&-\frac{1}{2}\nabla_\xi\mathcal{T}^{\eta\lambda}\big(f_\mathcal{T}g_{\eta\lambda}+f_\mathcal{Q}\mathcal{R}_{\eta\lambda}\big)
+\nabla_\xi\big(\mathbf{L}_mf_\mathcal{T}\big)-8\pi\nabla^\beta
\mathrm{E}_{\beta\xi}\bigg].
\end{align}

An important phenomenon in the framework of structural evolvement of
massive self-gravitating systems is supposed to be the anisotropy
which emerges due to the difference of pressure components in
tangential and radial direction. The anisotropy is found to be the
most likely element in the interior of compact stars, thus become a
significant topic of discussion for astrophysicists now a days.
Therefore, we consider the spacetime geometry associated with the
anisotropic fluid in its interior as
\begin{equation}\label{g5}
\mathcal{T}_{\beta\xi}=\big(\mu+P_\bot\big)\mathrm{K}_{\beta}\mathrm{K}_{\xi}
+P_\bot\left(g_{\beta\xi}-\mathrm{W}_\beta\mathrm{W}_\xi\right)+P_r\mathrm{W}_\beta\mathrm{W}_\xi,
\end{equation}
where the matter sector such as the radial/tangential pressure and
energy density are indicated by $P_r,~P_\bot$ and $\mu$,
respectively. Also, $\mathrm{W}_{\beta}$ is the four-vector and
$\mathrm{K}_{\beta}$ defines the four-velocity. The modified
gravitational field equations have the following trace given by
\begin{align}\nonumber
&3\nabla^{\beta}\nabla_{\beta}
f_\mathcal{R}-\mathcal{R}\left(\frac{\mathcal{T}}{2}f_\mathcal{Q}-f_\mathcal{R}\right)-\mathcal{T}(8\pi+f_\mathcal{T})+\frac{1}{2}
\nabla^{\beta}\nabla_{\beta}(f_\mathcal{Q}\mathcal{T})\\\nonumber
&+\nabla_\beta\nabla_\xi(f_\mathcal{Q}\mathcal{T}^{\beta\xi})-2f+(\mathcal{R}f_\mathcal{Q}+4f_\mathcal{T})\mathbf{L}_m
+2\mathcal{R}_{\beta\xi}\mathcal{T}^{\beta\xi}f_\mathcal{Q}\\\nonumber
&-2g^{\beta\xi} \frac{\partial^2\mathbf{L}_m}{\partial
g^{\beta\xi}\partial
g^{\eta\lambda}}\left(f_\mathcal{T}g^{\eta\lambda}+f_\mathcal{Q}\mathcal{R}^{\eta\lambda}\right)=0.
\end{align}
By substituting $f_\mathcal{Q}=0$ and the vacuum case in the above
field equations, the results of $f(\mathcal{R},\mathcal{T})$ and
$f(\mathcal{R})$ theories can be deduced, respectively. The
$\mathbb{EMT}$ affiliated with an electromagnetic field is
\begin{equation*}
\mathrm{E}_{\beta\xi}=\frac{1}{4\pi}\left[\frac{1}{4}g_{\beta\xi}\mathcal{Z}^{\eta\lambda}\mathcal{Z}_{\eta\lambda}
-\mathcal{Z}^{\eta}_{\beta}\mathcal{Z}_{\eta\xi}\right],
\end{equation*}
and the compact form of Maxwell equations is provided by
\begin{equation}\label{g5a}
\mathcal{Z}^{\beta\xi}_{;\xi}=4\pi \mathrm{J}^{\beta}, \quad
\mathcal{Z}_{[\beta\xi;\eta]}=0,
\end{equation}
where the current density $\mathrm{J}^{\beta}$ is defined in terms
of the charge density $\varpi_0$ as $\mathrm{J}^{\beta}=\varpi_0
\mathrm{K}^{\beta}$.

We suppose a static spherical line element in the following that
represents the interior of a self-gravitating system as
\begin{equation}\label{g6}
ds^2=-e^{\nu_1} dt^2+e^{\nu_2}
dr^2+r^2d\vartheta^2+r^2\sin^2\vartheta d\varphi^2,
\end{equation}
where $\nu_1=\nu_1(r)$ and $\nu_2=\nu_2(r)$. The first of
Eq.\eqref{g5a} takes the form
\begin{equation}
\tau''+\frac{1}{2r}\big[4-r(\nu_1'+\nu_2')\big]\tau'=4\pi\varpi_0
e^{\frac{\nu_1}{2}+\nu_2},
\end{equation}
giving rise to
\begin{equation}\label{g6a}
\tau'=\frac{\bar{s}}{r^2}e^{\frac{\nu_1+\nu_2}{2}},
\end{equation}
where $'=\frac{\partial}{\partial r}$ and $\bar{s}$ expresses the
total interior charge. Equation \eqref{g6a} produces the matter
Lagrangian as $\mathbf{L}_{m}=\frac{\bar{s}^2}{2r^4}$. Also, the
four quantities become for the metric \eqref{g6} as
\begin{equation}\label{g7}
\mathrm{W}_\beta=\delta_\beta^1 e^{\frac{\nu_2}{2}}, \quad
\mathrm{K}_\beta=-\delta_\beta^0 e^{\frac{\nu_1}{2}},
\end{equation}
with $\mathrm{K}^\beta\mathrm{K}_{\beta}=-1$ and $\mathrm{W}^\beta
\mathrm{K}_{\beta}=0$.

We are required to adopt a standard model of this theory to make our
results advantageous. Haghani \emph{et al.} \cite{22} discussed
cosmological applications of three different models in this
framework provided in the following
\begin{align}\label{g61}
f(\mathcal{R},\mathcal{T},\mathcal{R}_{\beta\xi}\mathcal{T}^{\beta\xi})&=
\mathcal{R}+\nu_3\mathcal{R}_{\beta\xi}\mathcal{T}^{\beta\xi},\\\label{g61a}
f(\mathcal{R},\mathcal{T},\mathcal{R}_{\beta\xi}\mathcal{T}^{\beta\xi})&=
\mathcal{R}\big(1+\nu_3\mathcal{R}_{\beta\xi}\mathcal{T}^{\beta\xi}\big),\\\label{g61b}
f(\mathcal{R},\mathcal{T},\mathcal{R}_{\beta\xi}\mathcal{T}^{\beta\xi})&=
\mathcal{R}+\nu_4\sqrt{\mid\mathcal{T}\mid}+\nu_3\mathcal{R}_{\beta\xi}\mathcal{T}^{\beta\xi},
\end{align}
where $\nu_3$ and $\nu_4$ treat as real-valued constants. They
analyzed the cosmic evolution and its dynamics for these models
with/without energy conservation. We observe that the functional of
this theory possesses the strong non-minimal matter geometry
coupling that leads to much lengthy and complex calculations. Hence,
we adopt the simplest standard model \eqref{g61} for our ease to
analyze the impact of such non-minimal coupling on considered
celestial structures. In this case, the resulting solution has an
oscillatory profile possessing alternating expanding and collapsing
phases for $\nu_3>0$. On the other hand, if $\nu_3<0$, the cosmic
scale factor has a hyperbolic cosine-type dependence. This model has
frequently been employed to study isotropic/anisotropic compact
stars with the help of different strategies and some admissible
values of $\nu_3$ are deduced \cite{22a,22b,25a}. The last term of
the above model comes out to be
\begin{eqnarray}\nonumber
\mathcal{Q}&=&\frac{1}{e^{\nu_2}}\bigg[\frac{\mu}{4}\left(2\nu_1''+\nu_1'^2-\nu_1'\nu_2'+\frac{4\nu_1'}{r}\right)
+\frac{P_r}{4}\left(\nu_1'\nu_2'-\nu_1'^2-2\nu_1''-\frac{4\nu_2'}{r}\right)\\\nonumber
&-&P_\bot\left(\frac{\nu_1'}{r}-\frac{\nu_2'}{r}-\frac{2e^{\nu_2}}{r^2}+\frac{2}{r^2}\right)\bigg].
\end{eqnarray}
The field equations \eqref{g2} and $\mathbb{EMT}$ \eqref{g4a} become
for the model \eqref{g61}, respectively, as
\begin{align}\nonumber
\mathcal{G}_{\beta\xi}&=\frac{\nu_3}{1-\frac{\nu_3 \bar{s}^2}{2r^4}}
\bigg[\left(\frac{8\pi}{\nu_3}+\frac{1}{2}\mathcal{R}\right)\mathcal{T}_{\beta\xi}
+\frac{8\pi}{\nu_3}\mathrm{E}_{\beta\xi}+\frac{1}{2}\left\{\mathcal{Q}
-\nabla_{\eta}\nabla_{\lambda}\mathcal{T}^{\eta\lambda}\right\}g_{\beta\xi}\\\label{g7a}
&-2\mathcal{R}_{\eta(\beta}\mathcal{T}_{\xi)}^{\eta}-\frac{1}{2}\Box\mathcal{T}_{\beta\xi}
+\nabla_{\eta}\nabla_{(\beta}\mathcal{T}_{\xi)}^{\lambda}
-\mathcal{R}^{\eta\lambda}\mathcal{Z}_{\beta\eta}\mathcal{Z}_{\xi\lambda}\bigg],\\\nonumber
\nabla^\beta
\mathcal{T}_{\beta\xi}&=\frac{2\nu_3}{\nu_3\mathcal{R}+16\pi}
\bigg[\nabla_\beta(\mathcal{R}^{\eta\beta}\mathcal{T}_{\eta\xi})-\frac{1}{2}
\mathcal{R}_{\eta\lambda}\nabla_\xi\mathcal{T}^{\eta\lambda}-\frac{1}{2}
\mathcal{T}_{\beta\xi}\nabla^\beta\mathcal{R}\\\label{g7b}
&-8\pi\nabla^\beta\mathrm{E}_{\beta\xi}-\mathcal{G}_{\beta\xi}\nabla^\beta\big(\mathbf{L}_m\big)\bigg].
\end{align}
Equation \eqref{g7a} provides the non-vanishing components of the
field equations as
\begin{align}\nonumber
8\pi\mu&=\frac{1}{e^{\nu_2}}\bigg[\frac{\nu_2'}{r}+\frac{e^{\nu_2}}{r^2}-\frac{1}{r^2}
+\nu_3\bigg\{\mu\bigg(\frac{3\nu_1'\nu_2'}{8}-\frac{\nu_1'^2}{8}
+\frac{\nu_2'}{r}+\frac{e^\nu_2}{r^2}-\frac{3\nu_1''}{4}-\frac{3\nu_1'}{2r}\\\nonumber
&-\frac{1}{r^2}\bigg)-\mu'\bigg(\frac{\nu_2'}{4}-\frac{1}{r}-\nu_1'\bigg)
+\frac{\mu''}{2}+P_r\bigg(\frac{\nu_1'\nu_2'}{8}
-\frac{\nu_1'^2}{8}-\frac{\nu_1''}{4}+\frac{\nu_2'}{2r}+\frac{\nu_2''}{2}\\\nonumber
&-\frac{3\nu_2'^2}{4}\bigg)+\frac{5\nu_2'P'_r}{4}-\frac{P''_r}{2}
+P_\bot\bigg(\frac{\nu_2'}{2r}-\frac{\nu_1'}{2r}+\frac{3e^{\nu_2}}{r^2}
-\frac{1}{r^2}\bigg)-\frac{P'_\bot}{r}+\frac{\bar{s}^2}{r^4}\\\label{g8}
&\times\bigg(\frac{\nu_2'}{2r}-\frac{e^{\nu_2}}{2r^2}+\frac{1}{2r^2}+\frac{\nu_1'\nu_2'}{8}
-\frac{\nu_1'^2}{8}-\frac{\nu_1''}{4}-\frac{e^{\nu_2}}{\nu_3}\bigg)\bigg\}\bigg],\\\nonumber
8\pi
P_r&=\frac{1}{e^{\nu_2}}\bigg[\frac{\nu_1'}{r}-\frac{e^{\nu_2}}{r^2}+\frac{1}{r^2}
+\nu_3\bigg\{\mu\bigg(\frac{\nu_1'\nu_2'}{8}+\frac{\nu_1'^2}{8}
-\frac{\nu_1''}{4}-\frac{\nu_1'}{2r}\bigg)-\frac{\nu_1'\mu'}{4}-P_r\\\nonumber
&\times\bigg(\frac{5\nu_1'^2}{8}-\frac{7\nu_1'\nu_2'}{8}+\frac{5\nu_1''}{4}-\frac{7\nu_2'}{2r}+\frac{\nu_1'}{r}-\nu_2'^2
-\frac{e^{\nu_2}}{r^2}+\frac{1}{r^2}\bigg)+P'_r\\\nonumber
&\times\bigg(\frac{\nu_1'}{4}+\frac{1}{r}\bigg)-P_\bot\bigg(\frac{\nu_2'}{2r}-\frac{\nu_1'}{2r}+\frac{3e^{\nu_2}}{r^2}
-\frac{1}{r^2}\bigg)+\frac{P'_\bot}{r}+\frac{\bar{s}^2}{r^4}\bigg(\frac{\nu_1'}{2r}+\frac{e^{\nu_2}}{2r^2}\\\label{g8a}
&-\frac{1}{2r^2}+\frac{\nu_1''}{4}+\frac{\nu_1'^2}{8}-\frac{\nu_1'\nu_2'}{8}+\frac{e^{\nu_2}}{\nu_3}\bigg)\bigg\}\bigg],\\\nonumber
8\pi
P_\bot&=\frac{1}{e^{\nu_2}}\bigg[\frac{1}{2}\bigg(\nu_1''+\frac{\nu_1'^2}{2}-\frac{\nu_1'\nu_2'}{2}
-\frac{\nu_2'}{r}+\frac{\nu_1'}{r}\bigg)
+\nu_3\bigg\{\mu\bigg(\frac{\nu_1'^2}{8}+\frac{\nu_1'\nu_2'}{8}-\frac{\nu_1''}{4}-\frac{\nu_1'}{2r}\bigg)\\\nonumber
&-\frac{\mu'\nu_1'}{4}+P_r\bigg(\frac{\nu_1'^2}{8}+\frac{3\nu_2'^2}{4}-\frac{\nu_1'\nu_2'}{8}+\frac{\nu_1''}{4}-\frac{\nu_2'}{2r}
-\frac{\nu_2''}{2}\bigg)-\frac{5\nu_2'P'_r}{4}+\frac{P''_r}{2}\\\nonumber
&-P_\bot\bigg(\frac{\nu_1'^2}{4}-\frac{\nu_1'\nu_2'}{4}+\frac{\nu_1''}{2}-\frac{\nu_2'}{r}+\frac{\nu_1'}{r}\bigg)
-P'_\bot\bigg(\frac{\nu_2'}{4}-\frac{\nu_1'}{4}-\frac{3}{r}\bigg)+\frac{P''_\bot}{2}\\\label{g8b}
&+\frac{\bar{s}^2}{r^4}\bigg(\frac{\nu_1'\nu_2'}{8}-\frac{\nu_1'^2}{8}-\frac{\nu_1''}{4}
+\frac{\nu_2'}{4r}-\frac{\nu_1'}{4r}-\frac{e^{\nu_2}}{\nu_3}\bigg)\bigg\}\bigg].
\end{align}

The relation between the mass function of spherical body and
$g_{rr}$ metric component has been established by Misner and Sharp
\cite{41b} as
\begin{equation}\nonumber
\bar{m}(r)=\frac{r}{2}\big(1-g^{\beta\xi}r_{,\beta}r_{,\xi}\big),
\end{equation}
producing in the case of charged fluid
\begin{equation}\label{g12a}
\bar{m}(r)=\frac{r}{2}\left(1-e^{-\nu_2}+\frac{\bar{s}^2}{r^2}\right).
\end{equation}

The system \eqref{g8}-\eqref{g8b} is observed to be highly
non-linear in terms of geometric quantities and the state
determinants, and contains six unknowns such as
$\nu_1,~\nu_2,~P_r,~P_\bot,~\mu$ and $\bar{s}$. We need some
restraints on that account so that the system can be easily solved.
Since we aim to investigate physical characteristics of different
compact quark structures, the $\mathbf{MIT}$ bag model
$\mathbf{E}o\mathbf{S}$ is considered which characterizes the
interior fluid configuration of such bodies \cite{27a}. This
constraint involves a bag constant $(\mathrm{B})$ and is presented
to be
\begin{equation}\label{g14a}
\mu=3P_r+4\mathrm{B}.
\end{equation}
Multiple values of this constant have been determined for different
compact structures that are used to analyze their internal
configurations \cite{41f,41f1}. Joining the highly complicated
system \eqref{g8}-\eqref{g8b} together with $\mathbf{E}o\mathbf{S}$
\eqref{g14a}, we have the following form of the matter variables as
\begin{align}\nonumber
\mu&=\bigg[\nu_3\bigg(\frac{9\nu_1''}{8}-\frac{e^{\nu_2}}{r^2}+\frac{1}{r^2}-\frac{\nu_2''}{8}
-\frac{5\nu_1'\nu_2'}{8}-\frac{\nu_2'^2}{16}-\frac{7\nu_2'}{2r}+\frac{3\nu_1'^2}{16}+\frac{7\nu_1'}{4r}\bigg)\\\nonumber
&+8\pi e^{\nu_2}\bigg]^{-1}\bigg[\frac{3}{4}\bigg(1+\frac{\nu_3
\bar{s}^2}{2r^4}\bigg)\bigg(\frac{\nu_2'}{r}+\frac{\nu_1'}{r}\bigg)+\mathrm{B}\bigg\{8\pi
e^{\nu_2}-\nu_3\bigg(\frac{4\nu_2'}{r}-\frac{3\nu_1'^2}{4}\\\label{g14b}
&+\nu_1'\nu_2'-\frac{3\nu_1''}{2}+\frac{\nu_2''}{2}+\frac{\nu_2'^2}{4}-\frac{\nu_1'}{r}+\frac{e^{\nu_2}}{r^2}
-\frac{1}{r^2}\bigg)\bigg\}\bigg],\\\nonumber
P_r&=\bigg[\nu_3\bigg(\frac{9\nu_1''}{8}-\frac{e^{\nu_2}}{r^2}+\frac{1}{r^2}-\frac{\nu_2''}{8}-\frac{5\nu_1'\nu_2'}{8}
-\frac{\nu_2'^2}{16}-\frac{7\nu_2'}{2r}+\frac{3\nu_1'^2}{16}+\frac{7\nu_1'}{4r}\bigg)\\\nonumber
&+8\pi e^{\nu_2}\bigg]^{-1}\bigg[\frac{1}{4}\bigg(1+\frac{\nu_3
\bar{s}^2}{2r^4}\bigg)\bigg(\frac{\nu_2'}{r}+\frac{\nu_1'}{r}\bigg)-\mathrm{B}\bigg\{8\pi
e^{\nu_2}-\nu_3\bigg(\frac{\nu_1'\nu_2'}{2}+\frac{\nu_2'}{r}\\\label{g14c}
&-\frac{2\nu_1'}{r}+\frac{e^{\nu_2}}{r^2}-\nu_1''-\frac{1}{r^2}\bigg)\bigg\}\bigg],\\\nonumber
P_\bot&=\bigg[8\pi
e^{\nu_2}+\nu_3\bigg(\frac{1}{r^2}-\frac{2e^{\nu_2}}{r^2}+\frac{\nu_1'^2}{4}+\frac{\nu_1''}{2}-\frac{\nu_1'\nu_2'}{4}
+\frac{\nu_1'}{r}-\frac{\nu_2'}{r}\bigg)\bigg]^{-1}\bigg[\frac{\nu_1'}{2r}-\frac{\nu_2'}{2r}\\\nonumber
&+\frac{\nu_1'^2}{4}-\frac{\nu_1'\nu_2'}{4}+\frac{\nu_1''}{2}+\nu_3\bigg\{8\pi
e^{\nu_2}+\nu_3\bigg(\frac{9\nu_1''}{8}-\frac{e^{\nu_2}}{r^2}+\frac{1}{r^2}
-\frac{\nu_2''}{8}-\frac{5\nu_1'\nu_2'}{8}-\frac{\nu_2'^2}{16}\\\nonumber
&-\frac{7\nu_2'}{2r}+\frac{3\nu_1'^2}{16}+\frac{7\nu_1'}{4r}\bigg)\bigg\}^{-1}\bigg\{\frac{1}{8r}\bigg(1+\frac{\nu_3
s^2}{2r^4}\bigg)\bigg(2\nu_1'\nu_2'^2+\nu_1'^3-\nu_1''\nu_2'-\nu_1'\nu_1''\\\nonumber
&-\nu_2'\nu_2''-\nu_1'\nu_2''+\frac{3\nu_1'^2\nu_2'}{2}
-\frac{3\nu_1'^2}{r}+\frac{3\nu_2'^3}{2}-\frac{\nu_2'^2}{r}-\frac{4\nu_1'\nu_2'}{r}\bigg)+2\pi
e^{\nu_2}\mathrm{B}\bigg(\nu_1'\nu_2'\\\nonumber
&-2\nu_1''+2\nu_2''-3\nu_2'^2-\frac{2\nu_1'}{r}+\frac{2\nu_2'}{r}\bigg)
+\frac{\nu_3\mathrm{B}}{16}\bigg(10\nu_1''\nu_2''-5\nu_1'\nu_2'\nu_2''+11\nu_1'\nu_1''\nu_2'\\\nonumber
&-11\nu_1''\nu_2'^2-\nu_1'^2\nu_2''
-2\nu_1''\nu_1'^2-10\nu_1''^2-\frac{7\nu_1'^2\nu_2'^2}{2}
+\frac{\nu_1'^3\nu_2'}{2}-\frac{36\nu_1'\nu_2'^2}{r}-\frac{8\nu_1'^3}{r}\\\nonumber
&+\frac{11\nu_1'\nu_2'^3}{2}+\frac{16\nu_1'^2\nu_2'}{r}
+\frac{28\nu_1''\nu_2'}{r}-\frac{8\nu_2'\nu_2''}{r}+\frac{12\nu_2'^3}{r}+\frac{3\nu_1'^4}{2}
-\frac{8\nu_1'^2}{r^2}-\frac{8\nu_2''e^{\nu_2}}{r^2}\\\nonumber
&+\frac{8\nu_2''}{r^2}-\frac{20\nu_2'^2}{r^2}-\frac{24\nu_1'\nu_1''}{r}+\frac{52\nu_1'\nu_2'}{r^2}+\frac{10\nu_1'\nu_2''}{r}
-\frac{4e^{\nu_2}\nu_1'\nu_2'}{r^2}+\frac{8e^{\nu_2}\nu_1''}{r^2}-\frac{8\nu_1''}{r^2}\\\nonumber
&+\frac{12\nu_2'^2e^{\nu_2}}{r^2}-\frac{8\nu_1'}{r^3}
-\frac{8e^{\nu_2}\nu_2'}{r^3}+\frac{8\nu_2'}{r^3}+\frac{8e^{\nu_2}\nu_1'}{r^3}\bigg)\bigg\}\bigg]
+\frac{\nu_3\bar{s}^2}{4r^4e^{\nu_2}}\bigg(\frac{\nu_1'\nu_2'}{2}-\frac{\nu_1'^2}{2}-\nu_1''\\\label{g14d}
&+\frac{\nu_2'}{r}-\frac{\nu_1'}{r}-\frac{4e^{\nu_2}}{\nu_3}\bigg).
\end{align}
Several exact solutions to the field equations representing quark
structures have been formulated in Einstein's gravity as well as
modified framework with the help of $\mathbf{E}o\mathbf{S}$
\eqref{g14a} \cite{41fa,41fb}. We develop such a solution in this
modified scenario in the presence of charge whose influence on
physical attributes of the considered stars shall later be checked
through graphical analysis.

\section{Tolman IV Solution and Some Constraints on Spherical Boundary}

In this section, we consider metric functions of Tolman IV geometry
to reduce the unknown quantities and get an analytic solution in
$f(\mathcal{R},\mathcal{T},\mathcal{Q})$ framework. This spacetime
acquired much significance in the scientific community and has the
following form
\begin{equation}\label{g15}
e^{\nu_{1}}=\mathcal{A}_{2}\left(1+\frac{r^2}{\mathcal{A}_{1}}\right),\quad\quad
e^{\nu_{2}}=\frac{1+\frac{2r^2}{\mathcal{A}_{1}}}{\left(1-\frac{r^2}{\mathcal{A}_{3}}\right)
\left(1+\frac{r^2}{\mathcal{A}_{1}}\right)},
\end{equation}
where $\mathcal{A}_{1},~\mathcal{A}_{2}$ and $\mathcal{A}_{3}$ are
real-valued constants which shall be calculated via boundary
conditions. The $g_{tt}$ and $g_{rr}$ metric functions under
consideration must obey the acceptability criteria \cite{41j}. In
order to check the acceptability of $e^{\nu_{1}}$ and $e^{\nu_{2}}$
\eqref{g15}, we take their first and second derivatives with respect
to $r$ as
\begin{align}\nonumber
\nu_{1}'(r)&=\frac{2r}{\mathcal{A}_{1}+r^2}, \quad
\nu_{1}''(r)=\frac{2\big(\mathcal{A}_{1}-r^2\big)}{\big(\mathcal{A}_{1}+r^2\big)^2},\\\nonumber
\nu_{2}'(r)&=\frac{2r\big(\mathcal{A}_{1}\mathcal{A}_{3}+2\mathcal{A}_{1}r^2+\mathcal{A}_{1}^2+2r^4\big)}
{\mathcal{A}_{1}^2\mathcal{A}_{3}+r^2\big(3\mathcal{A}_{1}\mathcal{A}_{3}-\mathcal{A}_{1}^2-3\mathcal{A}_{1}r^2
+2\mathcal{A}_{3}r^2-2r^4\big)},\\\nonumber
\nu_{2}''(r)&=2\left[\mathcal{A}_{1}^2\mathcal{A}_{3}+r^2\big(3\mathcal{A}_{1}\mathcal{A}_{3}-\mathcal{A}_{1}^2-3\mathcal{A}_{1}r^2
+2\mathcal{A}_{3}r^2-2r^4\big)\right]^{-2}\bigg[\mathcal{A}_{1}^3\mathcal{A}_{3}^2\\\nonumber
&+\mathcal{A}_{1}^4\mathcal{A}_{3}+r^2\bigg(4\mathcal{A}_{1}^3\mathcal{A}_{3}+19\mathcal{A}_{1}^2\mathcal{A}_{3}r^2
+24\mathcal{A}_{1}\mathcal{A}_{3}r^4+7\mathcal{A}_{1}^3r^2+6\mathcal{A}_{1}r^6\\\nonumber
&+\mathcal{A}_{1}^4+20\mathcal{A}_{1}^2r^4-6\mathcal{A}_{1}\mathcal{A}_{3}^2r^2+4\mathcal{A}_{3}r^6+4r^8
-\frac{9\mathcal{A}_{1}^2\mathcal{A}_{3}^2}{2}\bigg)\bigg].
\end{align}
It is observed at the core of compact star (i.e., $r=0$) that
$\nu_{1}'(0)=\nu_{2}'(0)=0,~\nu_{1}''(0)>0$ and $\nu_{2}''(0)>0$
everywhere, hence their acceptability is verified. Equations
\eqref{g14b}-\eqref{g14d} in relation with with these constants are
provided in Appendix \textbf{A}.

An immensely valuable tool to figure out a complete structure of
massive self-gravitating objects is the junction conditions which
are determined through smooth matching of the inner and outer
metrics on the spherical boundary. For this, we need an outer
spactime whose fundamental properties (the presence/absence of the
charge, static/non-static, etc.) must match with the interior
spacetime. Since we assume a static charged spherical interior, the
exterior geometry is taken as the Reissner-Nordstr\"{o}m metric.
Recall that there is a difference between the boundary conditions of
$\mathbb{GR}$ and $f(\mathcal{R})$ theory because the higher-order
geometric terms are present in the later case \cite{41jaa,41jaaa}.
However, the term $\mathcal{R}_{\beta\xi}\mathcal{T}^{\beta\xi}$ in
the model \eqref{g61} has no contribution in the current framework.
Therefore, the exterior line element can be taken as same as that of
$\mathbb{GR}$. The exterior metric with the total charge
$\bar{\mathcal{S}}$ and mass $\bar{\mathcal{M}}$ is given by
\begin{equation}\label{g20}
ds^2=-\left(1-\frac{2\bar{\mathcal{M}}}{r}+\frac{\bar{\mathcal{S}}^2}{r^2}\right)dt^2
+\frac{dr^2}{\left(1-\frac{2\bar{\mathcal{M}}}{r}+\frac{\bar{\mathcal{S}}^2}{r^2}\right)}
+r^2d\vartheta^2+r^2\sin^2\vartheta d\varphi^2.
\end{equation}
The first fundamental form of Darmois boundary conditions admits the
continuity of $g_{tt},~g_{rr}$ and $g_{tt,r}$ inner and outer metric
components across the boundary $\big(\Sigma: r=\mathcal{H}\big)$,
giving rise to the following expressions
\begin{eqnarray}\label{g21}
g_{tt}&{_{=}^{\Sigma}}&e^{\nu_{1}(\mathcal{H})}=\mathcal{A}_{2}\left(1+\frac{\mathcal{H}^2}{\mathcal{A}_{1}}\right)
=1-\frac{2\bar{\mathcal{M}}}{\mathcal{H}}+\frac{\bar{\mathcal{S}}^2}{\mathcal{H}^2},\\\label{g21a}
g_{rr}&{_{=}^{\Sigma}}&e^{\nu_{2}(\mathcal{H})}=\frac{1+\frac{2\mathcal{H}^2}{\mathcal{A}_{1}}}{\left(1
-\frac{\mathcal{H}^2}{\mathcal{A}_{3}}\right)\left(1+\frac{\mathcal{H}^2}{\mathcal{A}_{1}}\right)}
=\bigg(1-\frac{2\bar{\mathcal{M}}}{\mathcal{H}}+\frac{\bar{\mathcal{S}}^2}{\mathcal{H}^2}\bigg)^{-1},\\\label{g22}
\frac{\partial g_{tt}}{\partial
r}&{_{=}^{\Sigma}}&\nu_{1}'(\mathcal{H})=\frac{2\mathcal{H}}{\mathcal{A}_{1}+\mathcal{H}^2}
=\frac{2\bar{\mathcal{M}}}{\mathcal{H}^2}-\frac{2\bar{\mathcal{S}}^2}{\mathcal{H}^3},
\end{eqnarray}
whose simultaneous solution yields
\begin{align}\label{g23}
\mathcal{A}_{1}&=\frac{\mathcal{H}^2\left(3\bar{\mathcal{M}}\mathcal{H}-\mathcal{H}^2-2\bar{\mathcal{S}}^2\right)}
{\bar{\mathcal{S}}^2-\bar{\mathcal{M}}\mathcal{H}},\\\label{g24}
\mathcal{A}_{2}&=\frac{\mathcal{H}^2+2\bar{\mathcal{S}}^2-3\bar{\mathcal{M}}\mathcal{H}}{\mathcal{H}^2},
\quad \mathcal{A}_{3}=\frac{\mathcal{H}^3}{\bar{\mathcal{M}}}.
\end{align}

An important attribute of the pressure in radial direction is that
it must vanish at the boundary. This condition along with
\eqref{g14c}, \eqref{g23} and \eqref{g24} determines the value of
the bag constant as
\begin{align}\nonumber
\mathrm{B}&=\frac{1}{4r^4}\big[\big(\mathcal{H}^2\big(\mathcal{H}(\mathcal{H}-3\bar{\mathcal{M}})
+2\bar{\mathcal{S}}^2\big)+2\mathcal{H}^2\big(\bar{\mathcal{M}}\mathcal{H}-\bar{\mathcal{S}}^2\big)\big)
\big\{\big(\bar{\mathcal{M}}\mathcal{H}-\bar{\mathcal{S}}^2\big)\\\nonumber
&\times3\mathcal{H}^2\big(\mathcal{H}(\mathcal{H}-3\bar{\mathcal{M}})+2\bar{\mathcal{S}}^2\big)\big(\nu_3\big(\mathcal{H}^3
-4\bar{\mathcal{M}}\mathcal{H}^2\big)+8\pi\mathcal{H}^2\mathcal{H}^3\big)-7\nu_3\bar{\mathcal{M}}\mathcal{H}^4\\\nonumber
&\times\big(\bar{\mathcal{M}}\mathcal{H}-\bar{\mathcal{S}}^2\big)^2+\mathcal{H}^4\big(\mathcal{H}(\mathcal{H}-3\bar{\mathcal{M}})
+2\bar{\mathcal{S}}^2\big)^2\big(8\pi\mathcal{H}^3-3\nu_3\bar{\mathcal{M}}\big)+\mathcal{H}^5\\\nonumber
&\times\big(\bar{\mathcal{M}}\mathcal{H}-\bar{\mathcal{S}}^2\big)^2\big(\nu_3+16\pi\mathcal{H}^2\big)\big\}\big]^{-1}
\big[\mathcal{H}^2\big(\nu_3\bar{\mathcal{S}}^2+2\mathcal{H}^4\big)\big(\bar{\mathcal{M}}
\big(2\bar{\mathcal{S}}^2+\mathcal{H}^2\\\nonumber
&-3\bar{\mathcal{M}}\mathcal{H}\big)+2\mathcal{H}\big(\bar{\mathcal{M}}\mathcal{H}-\bar{\mathcal{S}}^2\big)\big)
\big(\mathcal{H}^2\big(\bar{\mathcal{M}}\mathcal{H}-\bar{\mathcal{S}}^2\big)+\mathcal{H}^2
\big(\mathcal{H}(\mathcal{H}-3\bar{\mathcal{M}})\\\label{g26}
&\times+2\bar{\mathcal{S}}^2\big)\big)^2\big].
\end{align}

We determine the constant triplet
$\big(\mathcal{A}_{1},~\mathcal{A}_{2},~\mathcal{A}_{3}\big)$ and
the bag constant by using the experimental information like radii
and masses of four different strange quark stars \cite{41k} provided
in Table $\mathbf{1}$. Since $\nu_3$ is any real-valued constant, we
can adopt any positive/negative value to explore whether the
corresponding solution is physically relevant or not. For instance,
we take $\nu_3\pm3$ in the current scenario. All these candidates
are observed to be consistent with Buchdhal's suggested limit
\cite{42a}, i.e.,
$\frac{2\bar{\mathcal{M}}}{\mathcal{H}}<\frac{8}{9}$. This triplet
is presented in Tables $\mathbf{2}$ and $\mathbf{3}$ for two values
of the charge as $\bar{\mathcal{S}}=0.2$ and $0.9$, respectively.
Tables $\mathbf{4}-\mathbf{7}$ deliver the values of the density,
radial pressure and the bag constant for each star model for
$\nu_3=\pm3$ and $\big(\bar{\mathcal{S}}=0.2$, $0.9\big)$.
\begin{table}[H]
\scriptsize \centering \caption{Preliminary data of different star
models \cite{hh}} \label{Table1} \vspace{+0.07in}
\setlength{\tabcolsep}{0.95em}
\begin{tabular}{cccccc}
\hline\hline Star Models & SMC X-4 & SAX J 1808.4-3658 & Her X-I &
4U 1820-30
\\\hline $Mass(M_{\bigodot})$ & 1.29 & 0.9 & 0.85 & 1.58
\\\hline
$\mathcal{H}(km)$ & 8.83 & 7.95 & 8.1 & 9.3
\\\hline
$\bar{\mathcal{M}}/\mathcal{H}$ & 0.215 & 0.166 & 0.154 & 0.249  \\
\hline\hline
\end{tabular}
\end{table}
\begin{table}[H]
\scriptsize \centering \caption{Unknown triplet for different star
models corresponding to $\bar{\mathcal{S}}=0.2$} \label{Table2}
\vspace{+0.07in} \setlength{\tabcolsep}{0.95em}
\begin{tabular}{cccccc}
\hline\hline Star Models & SMC X-4 & SAX J 1808.4-3658 & Her X-I &
4U 1820-30
\\\hline $\mathcal{A}_{1} (km^{2})$ & 129.904 & 191.485 & 229.920 & 87.891
\\\hline $\mathcal{A}_{2}$ & 0.3568 & 0.5021 & 0.5384 &
0.2525
\\\hline
$\mathcal{A}_{3} (km^{2})$ & 363.181 & 379.932 & 425.323 & 347.436\\
\hline\hline
\end{tabular}
\end{table}
\begin{table}[H]
\scriptsize \centering \caption{Unknown triplet for different star
models corresponding to $\bar{\mathcal{S}}=0.9$} \label{Table3}
\vspace{+0.07in} \setlength{\tabcolsep}{0.95em}
\begin{tabular}{cccccc}
\hline\hline Star Models & SMC X-4 & SAX J 1808.4-3658 & Her X-I &
4U 1820-30
\\\hline  $\mathcal{A}_{1} (km^{2})$ & 143.717 & 216.699 & 259.786 & 97.556
\\\hline $\mathcal{A}_{2}$ & 0.3766 & 0.5264 & 0.5619 &
0.2703
\\\hline
$\mathcal{A}_{3} (km^{2})$ & 363.181 & 379.932 & 425.323 & 347.436\\
\hline\hline
\end{tabular}
\end{table}
\begin{table}[H]
\scriptsize \centering \caption{Bag constant and matter variables
corresponding to different star models for $\nu_3=3$ and
$\bar{\mathcal{S}}=0.2$} \label{Table4} \vspace{+0.07in}
\setlength{\tabcolsep}{0.95em}
\begin{tabular}{cccccc}
\hline\hline Star Models & SMC X-4 & SAX J 1808.4-3658 & Her X-I &
4U 1820-30
\\\hline $\mathrm{B} (km^{-2})$ & 0.00011935 & 0.00012556 & 0.00011455 &
0.00011469
\\\hline
$\mu_c (gm/cm^3)$ & 1.44$\times$10$^{15}$ & 1.15$\times$10$^{15}$ &
9.88$\times$10$^{14}$ & 2.14$\times$10$^{15}$
\\\hline
$P_{c} (dyne/cm^2)$ & 2.51$\times$10$^{35}$ & 1.52$\times$10$^{35}$
& 1.21$\times$10$^{35}$ & 4.62$\times$10$^{35}$
\\\hline
$\mu_s (gm/cm^3)$ & 6.21$\times$10$^{14}$ & 6.55$\times$10$^{14}$ &
6.01$\times$10$^{14}$ & 6.15$\times$10$^{14}$
\\\hline
$\varrho_s$ & 0.203 & 0.159 & 0.148 & 0.249
\\\hline
$z_s$ & 0.298 & 0.212 & 0.192 & 0.416  \\
\hline\hline
\end{tabular}
\end{table}
\begin{table}[H]
\scriptsize \centering \caption{Bag constant and matter variables
corresponding to different star models for $\nu_3=3$ and
$\bar{\mathcal{S}}=0.9$} \label{Table5} \vspace{+0.07in}
\setlength{\tabcolsep}{0.95em}
\begin{tabular}{cccccc}
\hline\hline Star Models & SMC X-4 & SAX J 1808.4-3658 & Her X-I &
4U 1820-30
\\\hline $\mathrm{B} (km^{-2})$ & 0.00011771 & 0.00012163 & 0.00011059 &
0.00011396
\\\hline
$\mu_c (gm/cm^3)$ & 1.34$\times$10$^{15}$ & 1.06$\times$10$^{15}$ &
9.09$\times$10$^{14}$ & 1.97$\times$10$^{15}$
\\\hline
$P_{c} (dyne/cm^2)$ & 2.23$\times$10$^{35}$ & 1.31$\times$10$^{35}$
& 1.03$\times$10$^{35}$ & 4.09$\times$10$^{35}$
\\\hline
$\mu_s (gm/cm^3)$ & 6.09$\times$10$^{14}$ & 6.31$\times$10$^{14}$ &
5.77$\times$10$^{14}$ & 6.01$\times$10$^{14}$
\\\hline
$\varrho_s$ & 0.187 & 0.141 & 0.130 & 0.232
\\\hline
$z_s$ & 0.267 & 0.179 & 0.162 & 0.371  \\
\hline\hline
\end{tabular}
\end{table}
\begin{table}[H]
\scriptsize \centering \caption{Bag constant and matter variables
corresponding to different star models for $\nu_3=-3$ and
$\bar{\mathcal{S}}=0.2$} \label{Table6} \vspace{+0.07in}
\setlength{\tabcolsep}{0.95em}
\begin{tabular}{cccccc}
\hline\hline Star Models & SMC X-4 & SAX J 1808.4-3658 & Her X-I &
4U 1820-30
\\\hline $\mathrm{B} (km^{-2})$ & 0.00011928 & 0.00012551 & 0.00011451 &
0.00011461
\\\hline
$\mu_c (gm/cm^3)$ & 1.41$\times$10$^{15}$ & 1.12$\times$10$^{15}$ &
9.61$\times$10$^{14}$ & 2.08$\times$10$^{15}$
\\\hline
$P_{c} (dyne/cm^2)$ & 2.39$\times$10$^{35}$ & 1.45$\times$10$^{35}$
& 1.14$\times$10$^{35}$ & 4.44$\times$10$^{35}$
\\\hline
$\mu_s (gm/cm^3)$ & 6.04$\times$10$^{14}$ & 6.38$\times$10$^{14}$ &
5.85$\times$10$^{14}$ & 5.96$\times$10$^{14}$
\\\hline
$\varrho_s$ & 0.199 & 0.157 & 0.146 & 0.243
\\\hline
$z_s$ & 0.287 & 0.205 & 0.187 & 0.394  \\
\hline\hline
\end{tabular}
\end{table}
\begin{table}[H]
\scriptsize \centering \caption{Bag constant and matter variables
corresponding to different star models for $\nu_3=-3$ and
$\bar{\mathcal{S}}=0.9$} \label{Table7} \vspace{+0.07in}
\setlength{\tabcolsep}{0.95em}
\begin{tabular}{cccccc}
\hline\hline Star Models & SMC X-4 & SAX J 1808.4-3658 & Her X-I &
4U 1820-30
\\\hline $\mathrm{B} (km^{-2})$ & 0.00011759 & 0.00012151 & 0.00011049 &
0.00011385
\\\hline
$\mu_c (gm/cm^3)$ & 1.31$\times$10$^{15}$ & 1.03$\times$10$^{15}$ &
8.85$\times$10$^{14}$ & 1.92$\times$10$^{15}$
\\\hline
$P_{c} (dyne/cm^2)$ & 2.13$\times$10$^{35}$ & 1.24$\times$10$^{35}$
& 9.66$\times$10$^{34}$ & 3.96$\times$10$^{35}$
\\\hline
$\mu_s (gm/cm^3)$ & 5.96$\times$10$^{14}$ & 6.19$\times$10$^{14}$ &
5.66$\times$10$^{14}$ & 5.88$\times$10$^{14}$
\\\hline
$\varrho_s$ & 0.182 & 0.137 & 0.126 & 0.223
\\\hline
$z_s$ & 0.253 & 0.173 & 0.156 & 0.352  \\
\hline\hline
\end{tabular}
\end{table}
\begin{table}[H]
\scriptsize \centering \caption{Bag constant in terms of $MeV/fm^3$
for different parametric values} \label{Table8} \vspace{+0.07in}
\setlength{\tabcolsep}{0.95em}
\begin{tabular}{cccccc}
\hline\hline Star Models & SMC X-4 & SAX J 1808.4-3658 & Her X-I &
4U 1820-30
\\\hline $\nu_3=3,~\bar{\mathcal{S}}=0.2$ & 90.19 & 94.88 & 86.56 &
86.66
\\\hline
$\nu_3=3,~\bar{\mathcal{S}}=0.9$ & 88.95 & 91.91 & 83.57 & 86.11
\\\hline
$\nu_3=-3,~\bar{\mathcal{S}}=0.2$ & 90.13 & 94.84 & 86.53 & 86.61
\\\hline
$\nu_3=-3,~\bar{\mathcal{S}}=0.9$ & 88.86 & 91.82 & 83.49 & 86.03\\
\hline\hline
\end{tabular}
\end{table}

A specific range of the bag constant \big($60-80~MeV/fm^3$\big) has
been predicted through experiments in which compact quark models
show stable behavior \cite{aaa,bbb}. We observe that the values
corresponding to different considered models in this extended theory
(Table \textbf{8}) are slightly higher than the predicted range.
Multiple attempts, in this regard, have been done by $\mathrm{RHIC}$
and $\mathrm{CERN-SPS}$, and they deduced that the bag model
depending on density may supply a bigger range of the constant
$\mathrm{B}$.

\section{Physical Analysis of Compact Models}

This section interprets several physical attributes of the charged
anisotropic structures through graphical representation so that the
effect of this gravitational theory can be analyzed. Since the model
parameter $\nu_3$ is an unrestricted constant, its different values
would be helpful to explore the effect of the modified gravity. For
this, we use two different values of $\nu_3$ and charge along with
the preliminary data \big(given in Tables $\mathbf{1}-\mathbf{3}$)
to observe the nature of the extended solution
\eqref{g8}-\eqref{g8b}.

We plot temporal/radial metric functions, energy bounds, anisotropy
and the mass function for all considered stars. Moreover, the
interior charge in the field equations is also treated as an
unknown. Thus, we take its suggested known form to lessen the number
of unknown terms \cite{hha,hhb}. The following lines must be
memorized to understand all plots provided in this paper
\begin{itemize}
\item All thick lines correspond to $\bar{\mathcal{S}}=0.2$.
\item All dotted lines correspond to $\bar{\mathcal{S}}=0.9$.
\item Blue (thick and dotted) lines correspond to a model 4U 1820-30.
\item Red (thick and dotted) lines correspond to a model SMC X-4.
\item Black (thick and dotted) lines correspond to a model SAX J 1808.4-3658.
\item Green (thick and dotted) lines correspond to a model Her X-I.
\end{itemize}
Figure \textbf{1} provides the non-singular and increasing trend of
$g_{tt}$ and $g_{rr}$ metric components everywhere. Hence, the
acceptability of Tolman IV spacetime \eqref{g15} is verified.
\begin{figure}\center
\epsfig{file=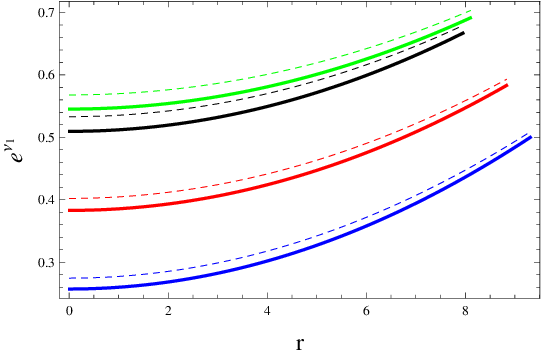,width=0.45\linewidth}\epsfig{file=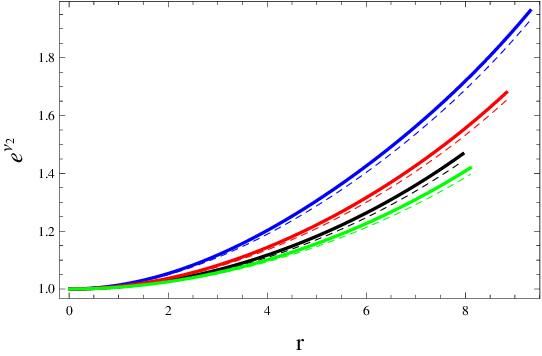,width=0.45\linewidth}
\caption{Metric coefficients corresponding to
$\bar{\mathcal{S}}=0.2$ (thick lines), $\bar{\mathcal{S}}=0.9$
(dotted lines), $\nu_3=-3$ (left plot) and $\nu_3=3$ (right plot)
for each star.}
\end{figure}

\subsection{Behavior of Matter Determinants}

An admissible behavior of the matter sector for
isotropic/anisotropic fluid distribution requires that these
parameters must be maximum and positive at $r=0$ and decreasing
outward. Figure \textbf{2} exhibits an acceptable behavior of the
density and radial/transverse components of pressure for chosen
values of $\nu_3$ and charge. All these parameters gain less values
in the interior of considered models for higher value of the charge.
It is also noted that $\nu_3=3$ produces denser interiors as
compared to its other adopted value. It is worthy to mention that
the radial pressure in this modified gravity corresponding to each
star disappears at the hypersurface. These compact systems become
densest for $\nu_3=3$ and $\bar{\mathcal{S}}=0.2$ among all the
adopted choices (Tables $\mathbf{4}-\mathbf{7}$). The regularity
conditions \big(such as $\frac{d\mu}{dr}|_{r=0} =
0,~\frac{dP_r}{dr}|_{r=0} = 0,~\frac{d^2\mu}{dr^2}|_{r=0} <
0$,~$\frac{d^2P_r}{dr^2}|_{r=0} < 0$\big) are also checked and
observed to be satisfied.
\begin{figure}\center
\epsfig{file=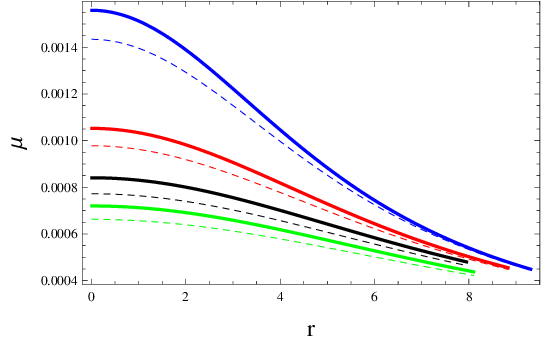,width=0.45\linewidth}\epsfig{file=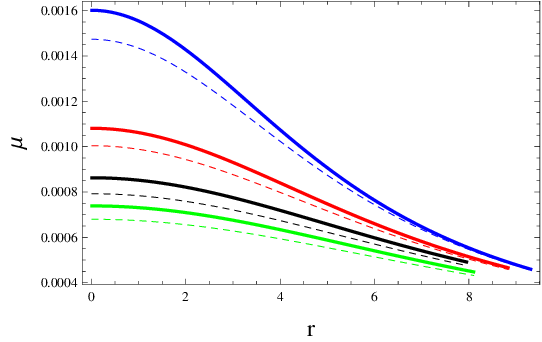,width=0.45\linewidth}
\epsfig{file=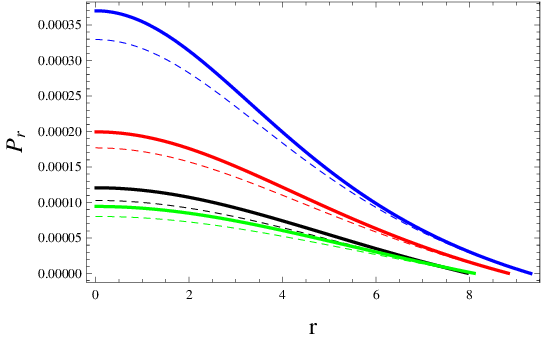,width=0.45\linewidth}\epsfig{file=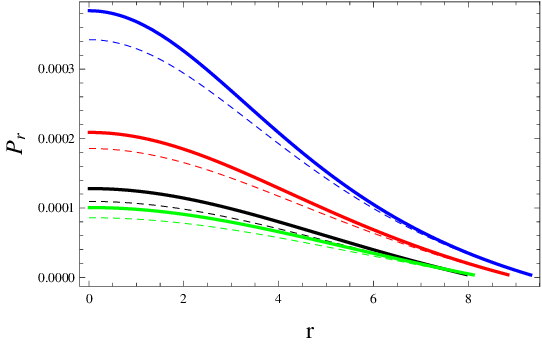,width=0.45\linewidth}
\epsfig{file=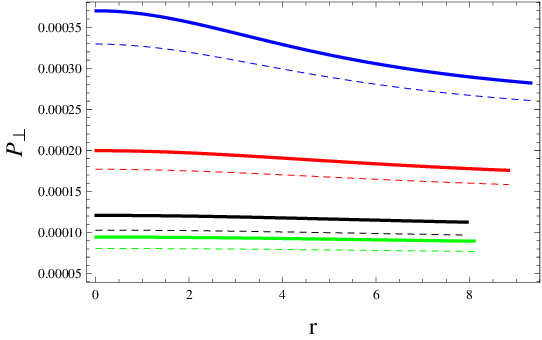,width=0.45\linewidth}\epsfig{file=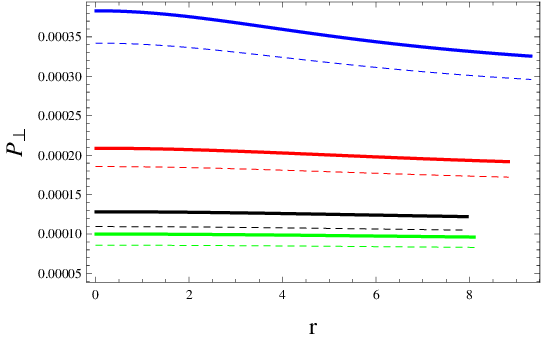,width=0.45\linewidth}
\caption{Matter determinants corresponding to
$\bar{\mathcal{S}}=0.2$ (thick lines), $\bar{\mathcal{S}}=0.9$
(dotted lines), $\nu_3=-3$ (left plots) and $\nu_3=3$ (right plots)
for each star.}
\end{figure}

\subsection{Pressure Anisotropy}

The anisotropy in the considered structures are defined by
$\Delta=P_\bot-P_r$ which can be calculated from Eqs.\eqref{g8a} and
\eqref{g8b}. Since the anisotropy has a great role in the evolution
of celestial systems, we check the impact of an electromagnetic
field on this factor. The anisotropy disappears at the star's center
and exhibits decreasing (inward) or increasing (outward) trend on
the basis whether the difference between both pressures is negative
or positive, respectively. This factor is observed to be null at the
core of each candidate and increase outward, as seen by Figure
$\mathbf{3}$. We also notice that the increment in charge causes
anisotropy to be reduced.

\subsection{Mass, Compactness and Surface Redshift}

A spherical structure \eqref{g6} has a mass interlinked with the
energy density given by
\begin{equation}\label{g32}
\bar{m}(r)=\frac{1}{2}\int_{0}^{\mathcal{H}}\breve{r}^2\mu
d\breve{r},
\end{equation}
where $\mu$ symbolizes the effective density in this modified theory
and given in Eq.\eqref{g14b}. Equations \eqref{g12a}, \eqref{g15},
\eqref{g23} and \eqref{g24} together yields, equivalently, as
\begin{equation}\label{g33}
\bar{m}(r)=\frac{r}{2}\left[1+\frac{\bar{s}^2}{r^2}-\frac{\big(\mathcal{H}^3-\bar{\mathcal{M}}r^2\big)
\big\{\mathcal{H}^3(\mathcal{H}-3\bar{\mathcal{M}})+\bar{\mathcal{M}}\mathcal{H}r^2
-\bar{\mathcal{S}}^2\big(r^2-2\mathcal{H}^2\big)\big\}}{\mathcal{H}^3\big\{\mathcal{H}^3(\mathcal{H}-3\bar{\mathcal{M}})
+2\bar{\mathcal{M}}\mathcal{H}r^2-2\bar{\mathcal{S}}^2\big(r^2-\mathcal{H}^2\big)\big\}}\right].
\end{equation}
Figure \textbf{3} represents the plots of the mass function for the
interior distribution of compact candidates. We observe that this
function possesses an increasing behavior outward. It is also shown
that the considered systems are more massive for the choices
$\nu_3=3$ and $\bar{\mathcal{S}}=0.2$. Moreover, the increasing
impact of charge provides less massive interiors.
\begin{figure}\center
\epsfig{file=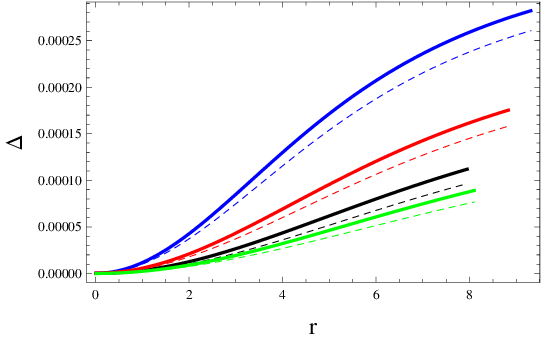,width=0.45\linewidth}\epsfig{file=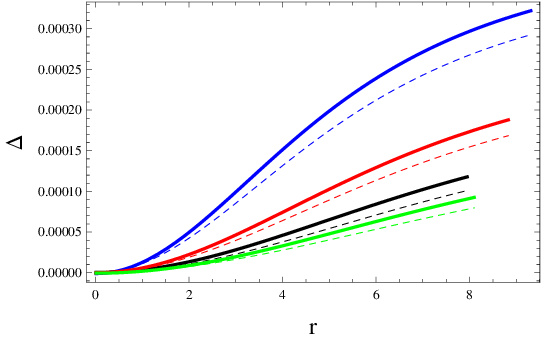,width=0.45\linewidth}
\epsfig{file=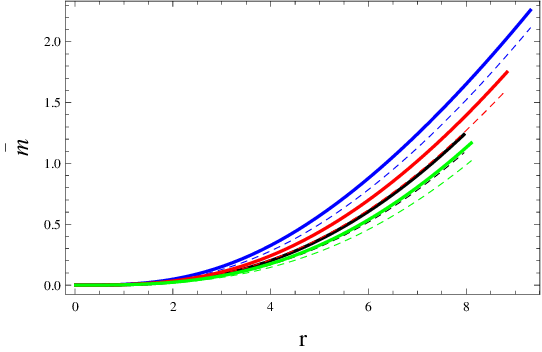,width=0.45\linewidth}\epsfig{file=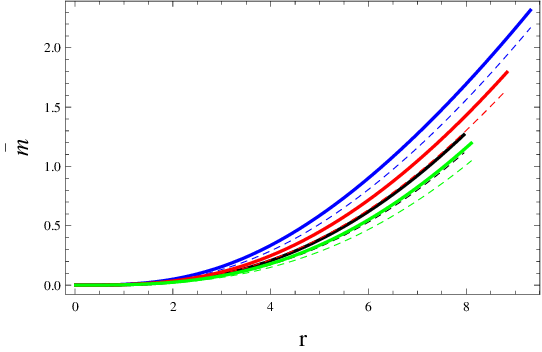,width=0.45\linewidth}
\caption{Anisotropy and mass corresponding to
$\bar{\mathcal{S}}=0.2$ (thick lines), $\bar{\mathcal{S}}=0.9$
(dotted lines), $\nu_3=-3$ (left plots) and $\nu_3=3$ (right plots)
for each star.}
\end{figure}

Some more physical parameters of compact stars must be analyzed
while discussing their evolution, such as the compactness and the
redshift. The former quantity (represented by $\varrho$) is defined
as a ratio of mass to radius of a self-gravitating body. It is given
by the following expression
\begin{align}\nonumber
\varrho(r)&=\frac{\bar{m}(r)}{r}\\\label{g34}
&=\frac{1}{2}\left[1+\frac{\bar{s}^2}{r^2}-\frac{\big(\mathcal{H}^3-\bar{\mathcal{M}}r^2\big)
\big\{\mathcal{H}^3(\mathcal{H}-3\bar{\mathcal{M}})+\bar{\mathcal{M}}\mathcal{H}r^2
-\bar{\mathcal{S}}^2\big(r^2-2\mathcal{H}^2\big)\big\}}{\mathcal{H}^3\big\{\mathcal{H}^3(\mathcal{H}-3\bar{\mathcal{M}})
+2\bar{\mathcal{M}}\mathcal{H}r^2-2\bar{\mathcal{S}}^2\big(r^2-\mathcal{H}^2\big)\big\}}\right].
\end{align}
A feasible spherically symmetric solution must have the value of
compactness less than $\frac{4}{9}$ everywhere in the interior
configuration \cite{42a}. The later parameter, mentioned above,
quantifies an increment in the electromagnetic waves (or radiations)
emitted by a heavily body when it undergoes specific reactions.
Mathematically, it is calculated as
\begin{equation}\label{g35}
z(r)=\frac{1-\sqrt{1-2\varrho(r)}}{\sqrt{1-2\varrho(r)}},
\end{equation}
leading to
\begin{equation}\label{g36}
z(r)=-1+\bigg[\frac{\big(\mathcal{H}^3-\bar{\mathcal{M}}r^2\big)\big\{\mathcal{H}^3(\mathcal{H}-3\bar{\mathcal{M}})
+\bar{\mathcal{M}}\mathcal{H}r^2-\bar{\mathcal{S}}^2\big(r^2-2\mathcal{H}^2\big)\big\}}{\mathcal{H}^3
\big\{\mathcal{H}^3(\mathcal{H}-3\bar{\mathcal{M}})+2\bar{\mathcal{M}}\mathcal{H}r^2
-2\bar{\mathcal{S}}^2\big(r^2-\mathcal{H}^2\big)\big\}}-\frac{\bar{s}^2}{r^2}\bigg]^{-\frac{1}{2}}.
\end{equation}
Buchdahl determined its maximum value at the hypersurface for
isotropic matter as $2$ which was later found to be $5.211$ for
anisotropic interior \cite{42b}. We plot both of these parameters in
Figure $\mathbf{4}$ for all choices of parameters and obtain
compatible behavior with the observational data. The increment in
charge and decrement in the bag constant decrease these entities.
Their values at $\Sigma: r=\mathcal{H}$ are provided in Tables
$\mathbf{4}-\mathbf{7}$ for all parametric values.
\begin{figure}\center
\epsfig{file=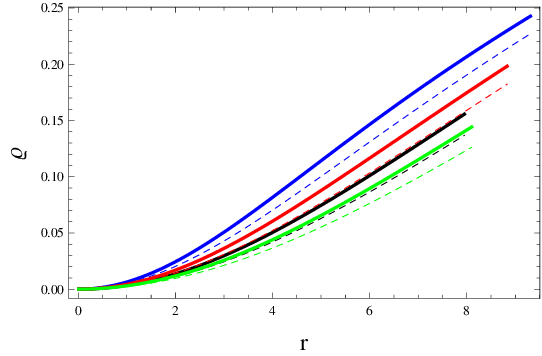,width=0.45\linewidth}\epsfig{file=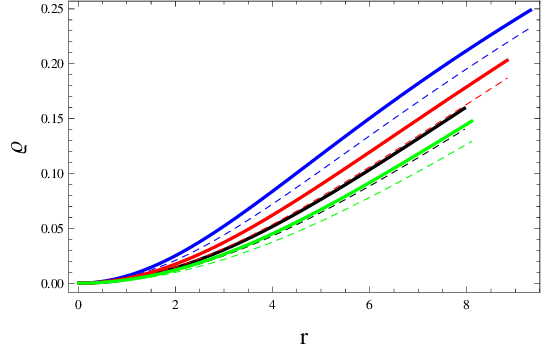,width=0.45\linewidth}
\epsfig{file=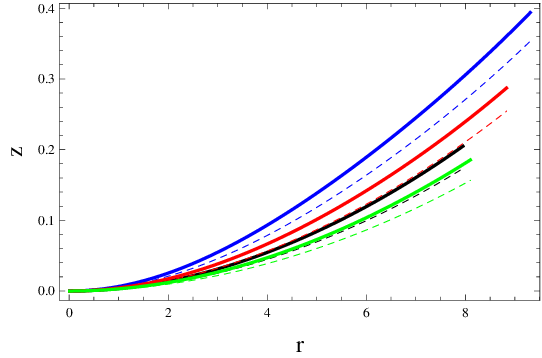,width=0.45\linewidth}\epsfig{file=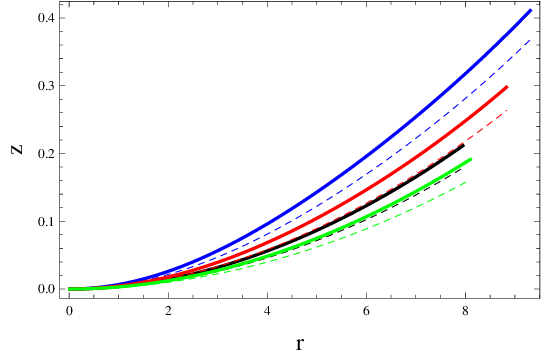,width=0.45\linewidth}
\caption{Compactness and redshift corresponding to
$\bar{\mathcal{S}}=0.2$ (thick lines), $\bar{\mathcal{S}}=0.9$
(dotted lines), $\nu_3=-3$ (left plots) and $\nu_3=3$ (right plots)
for each star.}
\end{figure}

\subsection{Energy Conditions}

In this subsection, we present some constraints through which the
existence of usual or exotic fluid in an interior geometry can be
confirmed. These bounds depend explicitly on state determinants and
are used extensively in the literature, titled as energy conditions.
Their satisfaction results in the presence of normal fluid in a
compact interior, thus leads to a viable model. Otherwise, there
must be an exotic matter inside a spherical geometry. They are given
as follows
\begin{itemize}
\item Dominant: $\mu-P_\bot \geq 0$, \quad $\mu-P_r+\frac{\bar{s}^2}{4\pi r^4} \geq 0$,
\item Strong: $\mu+2P_\bot+P_r+\frac{\bar{s}^2}{4\pi r^4} \geq 0$,
\item Weak: $\mu+\frac{\bar{s}^2}{8\pi r^4} \geq 0$, \quad $\mu+P_\bot+\frac{\bar{s}^2}{4\pi r^4} \geq 0$, \quad $\mu+P_r \geq 0$,
\item Null: $\mu+P_\bot+\frac{\bar{s}^2}{4\pi r^4} \geq 0$, \quad $\mu+P_r \geq 0$.
\end{itemize}
Figures \textbf{5} and \textbf{6} exhibit such bounds for $\nu_3=-3$
and $3$, respectively. We notice that all these conditions satisfy
everywhere for both values of charge, representing a viable extended
solution in $\mathcal{R}+\nu_3\mathcal{Q}$ gravity. It is worthy to
mention that we observe a contradiction of these results with those
of provided in \cite{22b}.
\begin{figure}\center
\epsfig{file=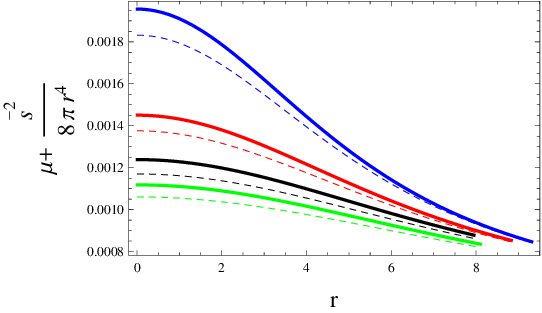,width=0.45\linewidth}\epsfig{file=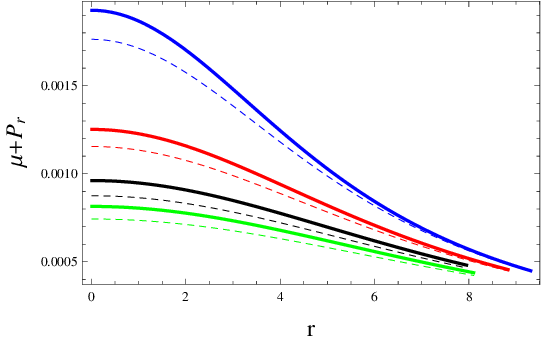,width=0.45\linewidth}
\epsfig{file=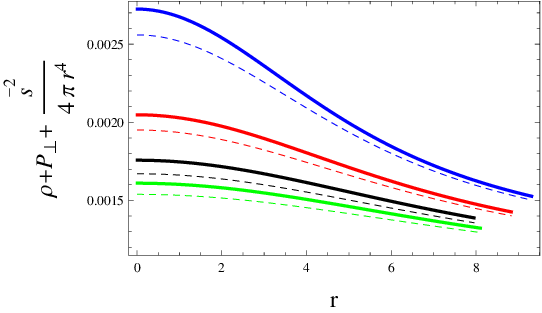,width=0.45\linewidth}\epsfig{file=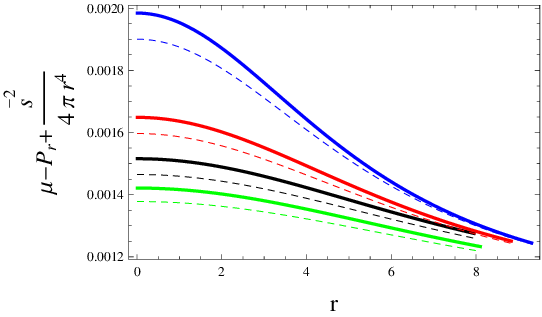,width=0.45\linewidth}
\epsfig{file=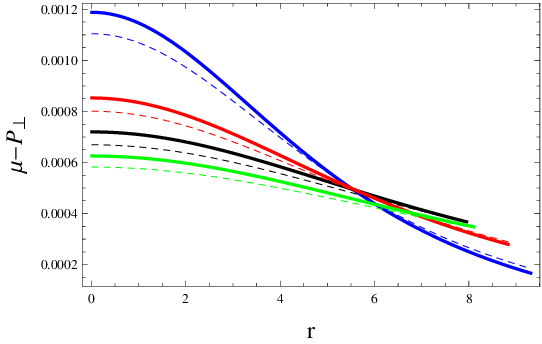,width=0.45\linewidth}\epsfig{file=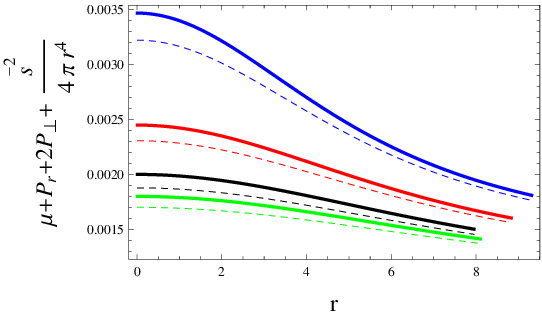,width=0.45\linewidth}
\caption{Energy conditions corresponding to $\nu_3=-3$ with
$\bar{\mathcal{S}}=0.2$ (thick lines) and $\bar{\mathcal{S}}=0.9$
(dotted lines) for each star.}
\end{figure}
\begin{figure}\center
\epsfig{file=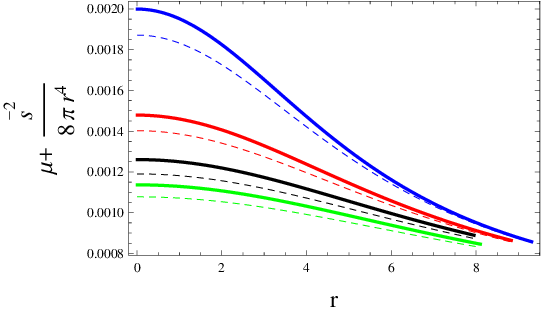,width=0.45\linewidth}\epsfig{file=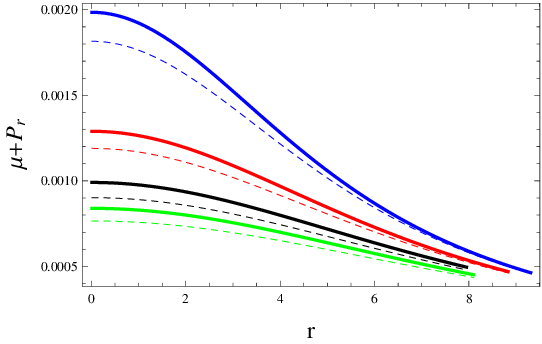,width=0.45\linewidth}
\epsfig{file=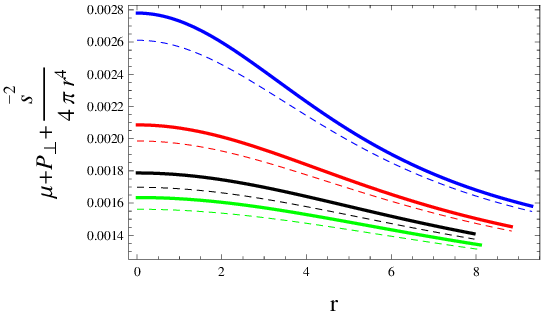,width=0.45\linewidth}\epsfig{file=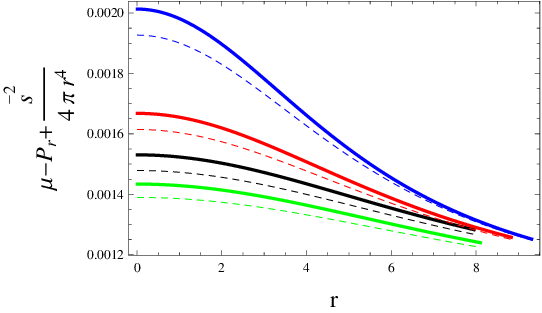,width=0.45\linewidth}
\epsfig{file=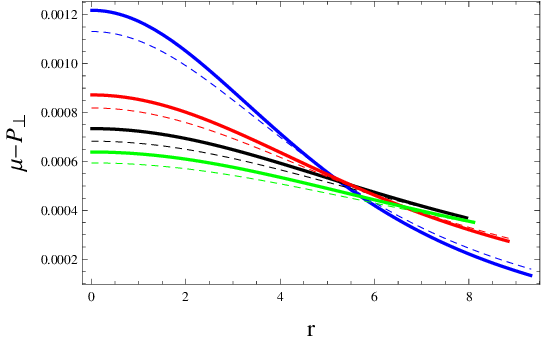,width=0.45\linewidth}\epsfig{file=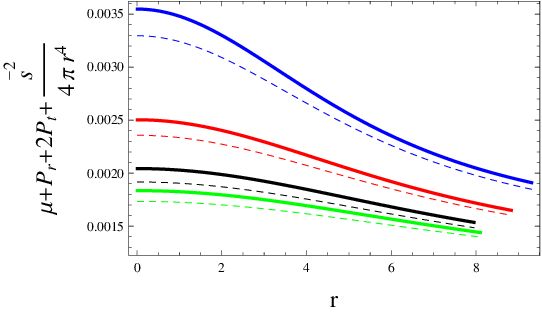,width=0.45\linewidth}
\caption{Energy conditions corresponding to $\nu_3=3$ with
$\bar{\mathcal{S}}=0.2$ (thick lines) and $\bar{\mathcal{S}}=0.9$
(dotted lines) for each star.}
\end{figure}

\subsection{Tolman-Opphenheimer-Volkoff Equation}

In this subsection, we plot different forces involving in the
Tolman-Opphenheimer-Volkoff ($\mathbb{TOV}$) equation corresponding
to this modified theory to check whether the developed model is in
stable equilibrium or not \cite{37ccc,37ddd}. We obtain the
following $\mathbb{TOV}$ equation from Eq.\eqref{g7b} given by
\begin{align}\nonumber
&\frac{dP_r}{dr}+\frac{\nu_1'}{2}\left(\mu+P_r\right)-\frac{2}{r}\left(P_\bot-P_r\right)
-\frac{2\nu_3}{e^{\nu_2}\big(\nu_3\mathcal{R}+16\pi\big)}\bigg[\frac{\nu_1'\mu}{8}\bigg(\nu_1'^2+2\nu_1''\\\nonumber
&-\nu_1'\nu_2'+\frac{4\nu_1'}{r}\bigg)-\frac{\mu'}{8}\bigg(\nu_1'^2-\nu_1'\nu_2'+2\nu_1''+\frac{4\nu_1'}{r}\bigg)
+P_r\bigg(\frac{5\nu_1'^2\nu_2'}{8}-\frac{5\nu_1'\nu_2'^2}{8}\\\nonumber
&-\frac{5\nu_2'^2}{2r}+\frac{7\nu_1''\nu_2'}{4}-\frac{\nu_1'''}{2}-\nu_1'\nu_1''+\frac{\nu_1'\nu_2''}{2}+\frac{2\nu_2''}{r}
+\frac{\nu_1'\nu_2'}{r}-\frac{\nu_2'}{r^2}-\frac{\nu_1''}{r}+\frac{\nu_1'}{r^2}\\\nonumber
&+\frac{2e^{\nu_2}}{r^3}-\frac{2}{r^3}\bigg)+\frac{P'_r}{8}\bigg(\nu_1'\nu_2'-2\nu_1''-\nu_1'^2+\frac{4\nu_2'}{r}\bigg)
+\frac{P_\bot}{r^2}\bigg(\nu_2'-\nu_1'+\frac{2e^{\nu_2}}{r}\\\nonumber
&-\frac{2}{r}\bigg)-\frac{P'_\bot}{r}\bigg(\frac{\nu_2'}{2}-\frac{\nu_1'}{2}
+\frac{e^{\nu_2}}{r}-\frac{1}{r}\bigg)-\left(\frac{\nu_1'}{r}-\frac{e^{\nu_2}}{r^2}+\frac{1}{r^2}
+\frac{2e^{\nu_2}}{\nu_3}\right)\\\label{g11}
&\times\bigg(\frac{\bar{s}\bar{s}'}{r^4}-\frac{2\bar{s}^2}{r^5}\bigg)\bigg]=0.
\end{align}

The compact notation of the above equation is
\begin{equation}\label{g36a}
f_h+f_a+f_g=0,
\end{equation}
where $f_h,~f_a$ and $f_g$ are hydrostatic, anisotropic and
gravitational, respectively, provided as
\begin{align}\nonumber
f_h&=-\frac{dP_r}{dr},\\\nonumber
f_a&=\frac{2}{r}\big(P_\bot-P_r\big),
\end{align}
and $f_g$ contains all remaining terms of Eq.\eqref{g11} with
opposite sign. Figure \textbf{7} indicates that the interiors of all
considered quark are in the hydrostatic equilibrium.

\subsection{Stability Analysis}

Stability plays a major role in studying the evolutionary patterns
of an astrophysical structure in our cosmos. The following lines
shall help us to study the stability of the considered modified
model \eqref{g61} with the help of three approaches.

\subsubsection{Causality Condition and Herrera Cracking Technique}

Since the fluid under consideration is anisotropic in nature, there
exist two sound speeds in tangential and radial directions that must
be less than the speed of light according to the causality condition
\cite{42d}, i.e.,
\begin{equation}\label{g51}
0 < v_{s\bot}^{2}=\frac{dP_{\bot}}{d\mu} < 1, \quad 0 <
v_{sr}^{2}=\frac{dP_{r}}{d\mu} < 1.
\end{equation}
We obtain $v_{sr}^{2}=\frac{1}{3}\in(0,1)$, thus do not need to plot
it. Furthermore, Figure \textbf{8} (upper plots) shows the
tangential sound speed. All candidates are appeared to be stable for
chosen parametric values except SAX J 1808.4-3658 which is unstable
only for $\nu_3=3$ and $\bar{\mathcal{S}}=0.9$.

The above two sound speeds are combined by Herrera \cite{42e} in a
single frame, known as cracking condition. According to this, a
stable interior can be obtained only if the following condition
holds
\begin{equation}\label{g52}
0 < \mid v_{s\bot}^{2}-v_{sr}^{2} \mid < 1.
\end{equation}
The lower plots of Figure \textbf{8} manifest that the considered
stars are stable for $\nu_3=-3$. However, two compact structures
such as SMC X-4 and SAX J 1808.4-3658 are unstable for
$\bar{\mathcal{S}}=0.9$ and both choices of charge along with
$\nu_3=3$, respectively.

\subsubsection{Adiabatic Index}

Another effective tool, in this regard, is the adiabatic index that
helps to check whether a system is stable or unstable. This
technique has been extensively discussed and used in the literature
\cite{42f}, and it was found that the value of this index
($\Gamma_{\mathrm{ad}}$) greater than $\frac{4}{3}$ leads to stable
models. Here, we define it as
\begin{equation}\label{g62}
\Gamma_{\mathrm{ad}}=\frac{P_{r}+\mu}{P_{r}}
\left(\frac{dP_{r}}{d\mu}\right)=\frac{P_{r}+\mu}{P_{r}}
\left(v_{sr}^{2}\right).
\end{equation}
Figure \textbf{9} depicts the graph of $\Gamma_{\mathrm{ad}}$ from
which stable compact models are achieved everywhere for all
parametric values.
\begin{figure}\center
\epsfig{file=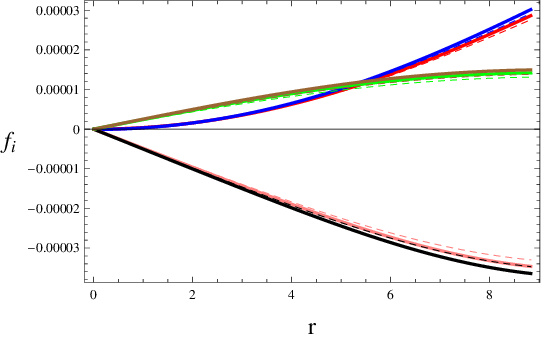,width=0.4\linewidth}\epsfig{file=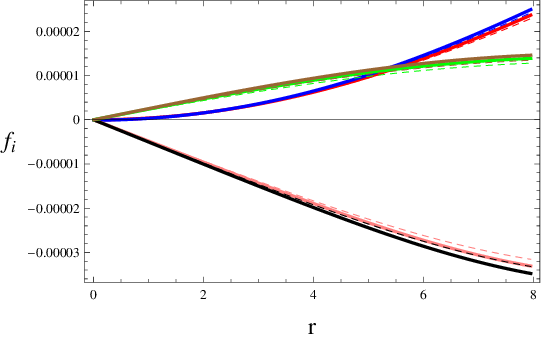,width=0.4\linewidth}
\epsfig{file=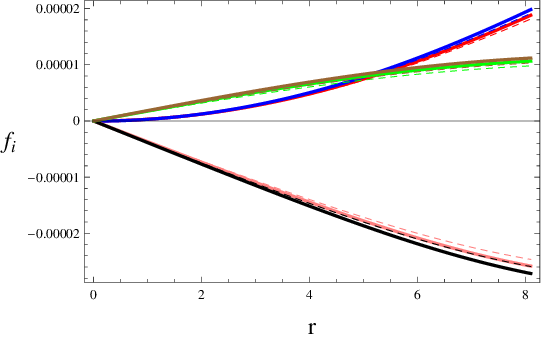,width=0.4\linewidth}\epsfig{file=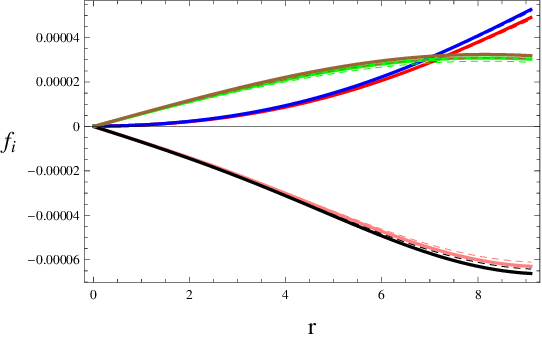,width=0.4\linewidth}\caption{Variation
in $f_{g}$ (black, pink), $f_{a}$ (red, blue) and $f_{h}$ (brown,
green) corresponding to SMC X-4 (upper left), SAX J 1808.4-3658
(upper right), Her X-I (lower left) and 4U 1820-30 (lower right).}
\end{figure}
\begin{figure}\center
\epsfig{file=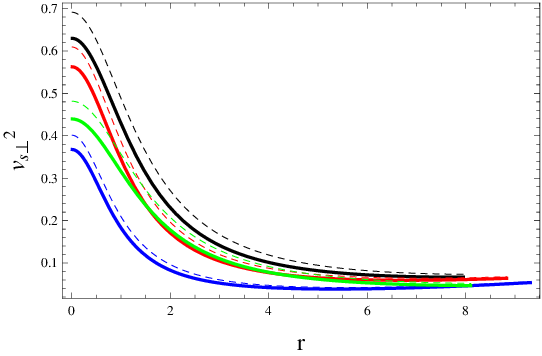,width=0.45\linewidth}\epsfig{file=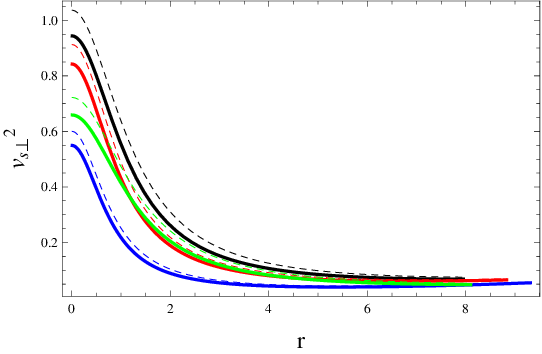,width=0.45\linewidth}
\epsfig{file=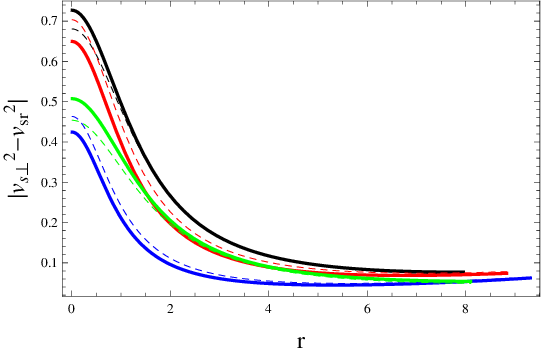,width=0.45\linewidth}\epsfig{file=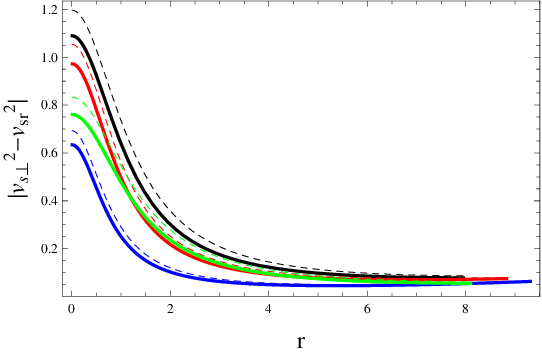,width=0.45\linewidth}
\caption{Tangential sound speed and $\mid v_{s\bot}^{2}-v_{sr}^{2}
\mid$ corresponding to $\bar{\mathcal{S}}=0.2$ (thick lines),
$\bar{\mathcal{S}}=0.9$ (dotted lines), $\nu_3=-3$ (left plots) and
$\nu_3=3$ (right plots) for each star.}
\end{figure}
\begin{figure}\center
\epsfig{file=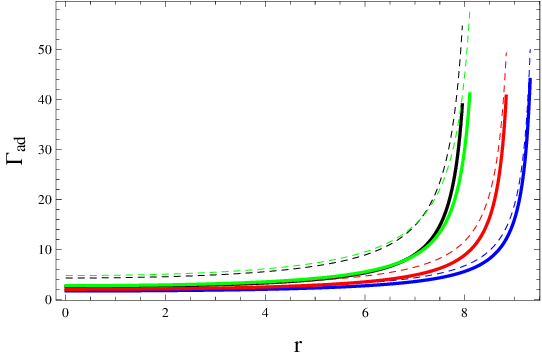,width=0.45\linewidth}\epsfig{file=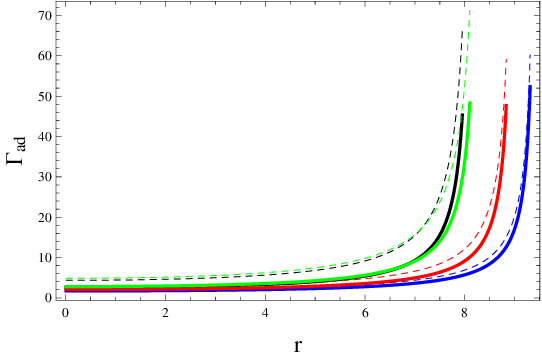,width=0.45\linewidth}
\caption{Adiabatic index corresponding to $\bar{\mathcal{S}}=0.2$
(thick lines), $\bar{\mathcal{S}}=0.9$ (dotted lines), $\nu_3=-3$
(left plot) and $\nu_3=3$ (right plot) for each star.}
\end{figure}

\section{Conclusions}

This paper discusses the impact of an electromagnetic field and the
non-minimal matter-geometry coupling on different quark star
candidates through a linear model $\mathcal{R}+\nu_3\mathcal{Q}$ by
choosing the coupling constant as $\nu_3=\pm3$. However, the
question arises in ones mind that why we have taken such large
values of the model parameter. Since we aimed to show that how
physical properties in the compact interiors behave by varying the
coupling parameter in this modified theory, those parametric values
must be taken which show the desired difference. In this regard, we
have initially done the whole analysis for small positive/negative
values of $\nu_3$, and found same profile of physical
characteristics everywhere for both considered values. However, we
have not included their plots in the paper. The matter Lagrangian
density for the charged fluid was taken as
$\mathbf{L}_{m}=-\frac{1}{4}\mathcal{Z}_{\beta\xi}\mathcal{Z}^{\beta\xi}$
\cite{22}, leading to $\mathbf{L}_m=\frac{\bar{s}^2}{2r^4}$. The
field equations and hydrostatic equilibrium condition have been
formulated in this modified theory. We have chosen $g_{tt}$ and
$g_{rr}$ metric functions of Tolman IV spacetime, and a particular
expression for the charge to reduce the number of unknowns in the
system of equations \eqref{g14b}-\eqref{g14d}. We have chosen four
star candidates such as SMC X-4,~SAX J 1808.4-3658,~Her X-I and 4U
1820-30 whose masses and radii are presented in Table \textbf{1}.
The unknown triplet ($\mathcal{A}_1,\mathcal{A}_2,\mathcal{A}_3$) in
Tolman IV spacetime has also been calculated at the spherical
boundary in terms of experimentally observed data (radii and masses)
of quark models. This triplet has been calculated in Tables
\textbf{2} and \textbf{3} for two different values of the charge as
0.2 and 0.9, respectively. We have plotted several physical
properties of self-gravitating systems for different choices of
coupling constant and charge. The developed state determinants are
maximum/minimum at the center/boundary for each star candidate,
leading to a physically acceptable model. The mass and anisotropy
have shown increasing behavior towards the spherical boundary for
all choices of parameters (Figure $\mathbf{3}$).

We have also noticed that the densest interiors of different
considered stars in this theory correspond to the positive choice of
$\nu_3$ along with less charge among all four discussed choices. The
trend of redshift and compactness has been detected as compatible.
The energy bounds are satisfied everywhere in the interior region of
each star, resulting in the viability of our developed model. We
have noticed that the increment in charge makes the considered
compact interiors less dense for both values of the model parameter
(Figure \textbf{2}). Moreover, the presence of charge produced less
anisotropic structures (Figure \textbf{3}). It must be mentioned
here that the interaction between gravity and an electromagnetic
field has been used to study the role of corresponding equations of
motion on the fields of purely physical nature \cite{37aggg}. The
central density, surface density as well as central radial pressure
have been evaluated in the interior of a compact candidate Her X-1
in the context of $\mathbb{GR}$ \cite{37ag}, from which we found
that this modified theory gains higher values in comparison with the
existing outcomes. We have also noticed the interior of Her X-1 in
this gravity to be the less dense as compared to that in
$f(\mathcal{G})$ theory \cite{38gg}. Further, the interior of the
quark star SAX J 1808.4-3658 has been explored in
$f(\mathcal{R},\mathcal{T})$ theory \cite{25aa} and found to be less
dense as opposed to $f(\mathcal{R},\mathcal{T},\mathcal{Q})$
framework. It is also observed that more suitable results are
obtained in
$f(\mathcal{R},\mathcal{T},\mathcal{R}_{\beta\xi}\mathcal{T}^{\beta\xi})$
theory for $\nu_3=-3$ as compared to \cite{25a,25b} and the solution
corresponding to $\nu_3=3$. Finally, stability has been observed
through different approaches. It is concluded that only two quark
stars such as Her X-I and 4U 1820-30 exhibit stable behavior for all
parametric choices, thus these results are consistent with
\cite{38}. However, the remaining models, namely SMC X-4 and SAX J
1808.4-3658 have shown stable behavior only for the negative value
of the coupling constant (Figure $\mathbf{8}$). On the other hand,
the compact star SMC X-4 was found to be stable for lower charge
with $\nu_3=3$, however, the increment in charge made it unstable.
It must be mentioned here that our results in this modified theory
reduce to $\mathbb{GR}$ for $\nu_3=0$.

\vspace{0.25cm}

\section*{Appendix A}
\renewcommand{\theequation}{A\arabic{equation}}
\setcounter{equation}{0} The matter sector \eqref{g14b}-\eqref{g14d}
in terms of Tolman IV spacetime is given by
\begin{align}\nonumber
\mu&=\frac{1}{r^4}\big[d_1^4 \big\{32 \pi d_3^2-d_3 \big(19 \nu_{3}
+32 \pi r^2\big)+16 \nu_{3}  r^2\big\}+2 d_1^3 \big\{2 d_3^2
\big(\nu_{3} +56 \pi r^2\big)\\\nonumber &-d_3 \big(112 \pi  r^4+69
\nu_{3} r^2\big)+58 \nu_{3} r^4\big\}+d_1 \big\{10 d_3^2\big(64 \pi
r^6+5 \nu_{3} r^4\big)-4 d_3 \\\nonumber &\times\big(160 \pi r^8+111
\nu_{3} r^6\big)+358 \nu_{3} r^8\big\}-4 r^6 \big\{d_3 \big(64 \pi
r^4+41 \nu_{3} r^2\big)-34 \nu_{3} r^4\\\nonumber &-4 d_3^2
\big(\nu_{3} +16 \pi r^2\big)\big\}+d_1^2 \big\{6 d_3^2 \big(96 \pi
r^4+5 \nu_{3} r^2\big)-d_3 \big(576 \pi r^6+383 \nu_{3}
r^4\big)\\\nonumber &+314 \nu_{3} r^6\big\}\big]^{-1}\big[d_1^4
\big\{32 \pi  \mathrm{B}  d_3^2 r^4+d_3 \big(-32 \pi \mathrm{B}
r^6+r^4 (6-40 \nu_{3}  \mathrm{B} )+3 \nu_{3}
\bar{s}^2\big)\\\nonumber &+r^6 (28 \nu_{3} \mathrm{B} -6)-3 \nu_{3}
r^2 \bar{s}^2\big\}-4 r^6 \big\{d_3 \big(64 \pi \mathrm{B}  r^8+r^6
(68 \nu_{3} \mathrm{B} +6)+3 \nu_{3}  r^2 \bar{s}^2\big)\\\nonumber
&-d_3^2 \big(64 \pi \mathrm{B} r^6+r^4 (4 \nu_{3}  \mathrm{B} +6)+3
\nu_{3} \bar{s}^2\big)-52 \nu_{3} \mathrm{B} r^8\big\}+2 d_1 r^4
\big\{d_3^2 \big(320 \pi \mathrm{B} r^6\\\nonumber &+r^4 (4 \nu_{3}
\mathrm{B} +30)+15 \nu_{3} \bar{s}^2\big)-4 d_3 \big(80 \pi
\mathrm{B}  r^8+r^6 (84 \nu_{3}  \mathrm{B} +6)+3 \nu_{3}  r^2
\bar{s}^2\big)\\\nonumber &+r^8 (260 \nu_{3} \mathrm{B} -6)-3
\nu_{3} r^4 \bar{s}^2\big\}+d_1^2 r^2 \big\{12 d_3^2 \big(48 \pi
\mathrm{B} r^6+r^4 (4-3 \nu_{3} \mathrm{B} )\\\nonumber &+2 \nu_{3}
\bar{s}^2\big)-d_3 \big(576 \pi \mathrm{B} r^8+2 r^6 (286 \nu_{3}
\mathrm{B} +9)+9 \nu_{3}  r^2 \bar{s}^2\big)+r^8 (452 \nu_{3}
\mathrm{B} -30)\\\nonumber &-15 \nu_{3} r^4 \bar{s}^2\big\}+2 d_1^3
\big\{d_3^2 \big(112 \pi \mathrm{B} r^6+r^4 (6-10 \nu_{3} \mathrm{B}
)+3 \nu_{3} \bar{s}^2\big)+d_3 \big(3 \nu_{3}  r^2
\bar{s}^2\\\label{A1} &-112 \pi \mathrm{B} r^8+r^6 (6-114 \nu_{3}
\mathrm{B} )\big)+4 r^8 (22 \nu_{3} \mathrm{B} -3)-6 \nu_{3} r^4
\bar{s}^2\big\}\big],\\\nonumber P_r&=\frac{1}{r^4}\big[d_1^4
\big\{32 \pi d_3^2-d_3 \big(19 \nu_{3} +32 \pi r^2\big)+16 \nu_{3}
r^2\big\}+2 d_1^3 \big\{2 d_3^2 \big(\nu_{3} +56 \pi
r^2\big)\\\nonumber &-d_3 \big(112 \pi  r^4+69 \nu_{3} r^2\big)+58
\nu_{3} r^4\big\}+d_1 \big\{10 d_3^2\big(64 \pi r^6+5 \nu_{3}
r^4\big)-4 d_3 \\\nonumber &\times\big(160 \pi r^8+111 \nu_{3}
r^6\big)+358 \nu_{3} r^8\big\}-4 r^6 \big\{d_3 \big(64 \pi r^4+41
\nu_{3} r^2\big)-34 \nu_{3} r^4\\\nonumber &-4 d_3^2 \big(\nu_{3}
+16 \pi r^2\big)\big\}+d_1^2 \big\{6 d_3^2 \big(96 \pi r^4+5 \nu_{3}
r^2\big)-d_3 \big(576 \pi r^6+383 \nu_{3} r^4\big)\\\nonumber &+314
\nu_{3} r^6\big\}\big]^{-1}\big[\big(d_1+2 r^2\big)
\big(r^2-d_3\big) \big\{d_1^3 \big(32 \pi \mathrm{B} d_3 r^4-2 r^4
(6 \nu_{3}  \mathrm{B} +1)\\\nonumber &-\nu_{3} \bar{s}^2\big)-2 r^4
\big(d_3 \big(-64 \pi \mathrm{B}  r^6+r^4 (2-4 \nu_{3}  \mathrm{B}
)+\nu_{3} \bar{s}^2\big)+28 \nu_{3} \mathrm{B} r^6\big)-2
d_1^2\\\nonumber &\times \big(d_3 \big(-80 \pi  \mathrm{B} r^6+r^4
(2-6 \nu_{3} \mathrm{B} )+\nu_{3} \bar{s}^2\big)+r^6 (36 \nu_{3}
\mathrm{B} +2)+\nu_{3} r^2 \bar{s}^2\big)\\\label{A2} &+d_1 \big(4
d_3 \big(64 \pi \mathrm{B} r^8+r^6 (7 \nu_{3} \mathrm{B} -2)-\nu_{3}
r^2 \bar{s}^2\big)-2 r^8 (62 \nu_{3} \mathrm{B} +1)-\nu_{3} r^4
\bar{s}^2\big)\big\}\big],\\\nonumber P_\bot&=\big[2 d_3 r^4
\big(d_1+2 r^2\big) \big\{d_1^2 \big(8 \pi  d_3-3 \nu_{3} \big)+d_1
r^2 \big(32 \pi  d_3-11 \nu_{3} \big)+2 \big(d_3 r^2\\\nonumber
&\times \big(\nu_{3} +16 \pi  r^2\big)-6 \nu_{3} r^4\big)\big\}
\big\{d_1^4 \big(d_3 \big(19 \nu_{3} +32 \pi r^2\big)-32 \pi
d_3^2-16 \nu_{3} r^2\big)-2 d_1^3\\\nonumber &\times \big(2 d_3^2
\big(\nu_{3} +56 \pi r^2\big)-d_3 \big(112 \pi r^4+69 \nu_{3}
r^2\big)+58 \nu_{3} r^4\big)+d_1 \big(d_3 \big(640 \pi
r^8\\\nonumber &+444 \nu_{3} r^6\big)-10 d_3^2 \big(64 \pi r^6+5
\nu_{3} r^4\big)-358 \nu_{3} r^8\big)+4 r^6 \big(d_3 \big(64 \pi
r^4+41 \nu_{3} r^2\big)\\\nonumber &-4 d_3^2 \big(\nu_{3} +16 \pi
r^2\big)-34 \nu_{3} r^4\big)+d_1^2 \big(d_3 \big(576 \pi r^6+383
\nu_{3} r^4\big)-314 \nu_{3} r^6\\\nonumber &-6 d_3^2 \big(96 \pi
r^4+5 \nu_{3} r^2\big)\big)\big\}\big]^{-1}\big[8 \big\{612
\bar{s}^2 \nu_{3} ^3 r^8-4 \nu_{3}  \big(2 (62 \nu_{3} \mathrm{B}
-51) r^6+12\\\nonumber &\times (17+58 \pi ) \bar{s}^2 \nu_{3}
r^2+261 \bar{s}^2 \nu_{3} ^2\big) d_3 r^6+\big(3072 \pi ^2 \bar{s}^2
\nu_{3} r^4+16 \pi \big(232 \bar{s}^2 \nu_{3} r^4\\\nonumber &+16 (4
\nu_{3} \mathrm{B} -3) r^8+265 \bar{s}^2 \nu_{3} ^2 r^2\big)+\nu_{3}
\big((472 \nu_{3} \mathrm{B} -644) r^6+1112 \bar{s}^2 \nu_{3}
r^2\\\nonumber &+475 \bar{s}^2 \nu_{3} ^2\big)\big) d_3^2
r^4-\big(4096 \pi ^2 \bar{s}^2 \big(r^2+\nu_{3} \big) r^4+16 \pi
\big(276 \bar{s}^2 \nu_{3} r^4+93 \bar{s}^2 \nu_{3} ^2
r^2\big)\\\nonumber &+16 (3 \nu_{3} \mathrm{B} -4) r^8+\nu_{3}
\big(4 (12 \nu_{3} \mathrm{B} -55) r^6+256 \bar{s}^2 \nu_{3}  r^2+77
\bar{s}^2 \nu_{3} ^2\big)\big) d_3^3 r^2\\\nonumber &+4 \big(256 \pi
^2 \bar{s}^2 \big(4 r^2+\nu_{3} \big) r^4+\nu_{3} \big(-4 r^6+4
\bar{s}^2 \nu_{3} r^2+\bar{s}^2 \nu_{3} ^2\big)-32 \pi \big(2
r^8\\\nonumber &-4 \bar{s}^2 \nu_{3} r^4-\bar{s}^2 \nu_{3} ^2
r^2\big)\big) d_3^4\big\} r^8-4 d_1 \big\{-4752 \bar{s}^2 \nu_{3} ^3
r^8+2 \nu_{3} \big(10 (222 \nu_{3} \mathrm{B} -161) \\\nonumber
&\times r^6+4 (827+2698 \pi ) \bar{s}^2 \nu_{3} r^2+3947 \bar{s}^2
\nu_{3} ^2\big) d_3 r^6-4 \big(5888 \pi ^2 \bar{s}^2 \nu_{3}
r^4+4\pi\\\nonumber &\times \big(16 (33 \nu_{3} \mathrm{B} -23)
r^8+1876 \bar{s}^2 \nu_{3} r^4+2013 \bar{s}^2 \nu_{3} ^2
r^2\big)+\nu_{3} \big( (430 \nu_{3} \mathrm{B} -419)
\\\nonumber &\times3r^6+2234 \bar{s}^2 \nu_{3} r^2+835
\bar{s}^2 \nu_{3} ^2\big)\big) d_3^2 r^4+\big(2048 \pi ^2 \bar{s}^2
\big(16 r^2+15 \nu_{3} \big) r^4+16 \pi\\\nonumber &\times \big(16
(29 \nu_{3} \mathrm{B} -30) r^8+2228 \bar{s}^2 \nu_{3} r^4+663
\bar{s}^2 \nu_{3} ^2 r^2\big)+\nu_{3} \big(4 (368 \nu_{3} \mathrm{B}
-401) r^6\\\nonumber &+1944 \bar{s}^2 \nu_{3} r^2+499 \bar{s}^2
\nu_{3} ^2\big)\big) d_3^3 r^2-\big(1024 \pi ^2 \bar{s}^2 \big(32
r^2+7 \nu_{3} \big) r^4+16 \pi \big(16 r^8\\\nonumber &\times(5
\nu_{3} \mathrm{B} -7)+244 \bar{s}^2 \nu_{3} r^4+53 \bar{s}^2
\nu_{3} ^2 r^2\big)+\nu_{3} \big((80 \nu_{3} \mathrm{B} -92) r^6+136
\bar{s}^2 \nu_{3} r^2\\\nonumber &+25 \bar{s}^2 \nu_{3} ^2\big)\big)
d_3^4\big\} r^6-2 d_1^2 \big\{-15067 \bar{s}^2 \nu_{3} ^3
r^8+\nu_{3} \big(8 (1911 \nu_{3} \mathrm{B} -1297) r^6+24\\\nonumber
&\times (919+2880 \pi ) \bar{s}^2 \nu_{3} r^2+23881 \bar{s}^2
\nu_{3} ^2\big) d_3 r^6-2 \big(37888 \pi ^2 \bar{s}^2 \nu_{3} r^4+16
\pi \big(296\\\nonumber &\times (3 \nu_{3} \mathrm{B} -2) r^8+3154
\bar{s}^2 \nu_{3} r^4+3117 \bar{s}^2 \nu_{3} ^2 r^2\big)+\nu_{3}
\big((9424 \nu_{3} \mathrm{B} -7924) r^6\\\nonumber &+14480
\bar{s}^2 \nu_{3} r^2+4359 \bar{s}^2 \nu_{3} ^2\big)\big) d_3^2
r^4+2 \big(1024 \pi ^2 \bar{s}^2 \big(54 r^2+47 \nu_{3} \big) r^4+16
\pi \big(16 \\\nonumber &\times(51 \nu_{3} \mathrm{B} -47) r^8+3704
\bar{s}^2 \nu_{3} r^4+915 \bar{s}^2 \nu_{3} ^2 r^2\big)+\nu_{3}
\big(20 (164 \nu_{3} \mathrm{B} -111)r^6\\\nonumber &+2632 \bar{s}^2
\nu_{3} r^2+501 \bar{s}^2 \nu_{3} ^2\big)\big) d_3^3 r^2-2 \big(2048
\pi ^2 \bar{s}^2 \big(27 r^2+5 \nu_{3} \big) r^4+\nu_{3}
\big(60\\\nonumber &\times (3 \nu_{3} \mathrm{B} -1) r^6+164
\bar{s}^2 \nu_{3} r^2+15 \bar{s}^2 \nu_{3} ^2\big)+32 \pi \big(20 (5
\nu_{3} \mathrm{B} -4) r^8+173 \bar{s}^2 \nu_{3} r^4\\\nonumber &+30
\bar{s}^2 \nu_{3} ^2 r^2\big)\big) d_3^4\big\} r^4+d_1^4 \big\{13150
\bar{s}^2 \nu_{3} ^3 r^6-8 \nu_{3} \big(3 (598 \nu_{3} \mathrm{B}
-379) r^6+4 (673\\\nonumber &+1922 \pi ) \bar{s}^2 \nu_{3} r^2+2328
\bar{s}^2 \nu_{3} ^2\big) d_3 r^4+\big(69120 \pi ^2 \bar{s}^2
\nu_{3} r^4+8 \pi \big((73 \nu_{3} \mathrm{B} -45) \\\nonumber
&\times48r^8+12548 \bar{s}^2 \nu_{3} r^4+10139 \bar{s}^2 \nu_{3} ^2
r^2\big)+\nu_{3} \big(4247 \bar{s}^2 \nu_{3} ^2+26288 \bar{s}^2
\nu_{3} r^2\\\nonumber &+2 (7724 \nu_{3} \mathrm{B} -6689)
r^6\big)\big) d_3^2 r^2+4 \big(640 \pi ^2 r^2 \big(44 r^2+5 \nu_{3}
\big) \bar{s}^2+\nu_{3} \big(34 \bar{s}^2 \nu_{3}\\\nonumber &
+(37-70 \nu_{3} \mathrm{B} ) r^4\big)+4 \pi \big(40 (7 \nu_{3}
\mathrm{B} -5) r^6+350 \bar{s}^2 \nu_{3} r^2+27 \bar{s}^2 \nu_{3}
^2\big)\big) d_3^4-2\\\nonumber &\times\big(5120 \pi ^2 \bar{s}^2
\big(11 r^2+8 \nu_{3} \big) r^4+4 \pi \big(16 (169 \nu_{3}
\mathrm{B} -160) r^8+14268 \bar{s}^2 \nu_{3} r^4\\\nonumber &+1995
\bar{s}^2 \nu_{3} ^2 r^2\big)+\nu_{3} \big(3 (734 \nu_{3} \mathrm{B}
-433) r^6+1166 \bar{s}^2 \nu_{3} r^2+90 \bar{s}^2 \nu_{3}
^2\big)\big) d_3^3\big\} r^2\\\nonumber &+d_1^3 \big\{25874
\bar{s}^2 \nu_{3} ^3 r^8-\nu_{3} \big(620 (44 \nu_{3} \mathrm{B}
-29) r^6+8 (4997+15002 \pi ) \bar{s}^2 \nu_{3} r^2\\\nonumber
&+38819 \bar{s}^2 \nu_{3} ^2\big) d_3 r^6+\big(133120 \pi ^2
\bar{s}^2 \nu_{3} r^4+16 \pi \big(11556 \bar{s}^2 \nu_{3} r^4+10379
\bar{s}^2 \nu_{3} ^2 r^2\\\nonumber &+16 (201 \nu_{3} \mathrm{B}
-130) r^8\big)+\nu_{3} \big(4 (8104 \nu_{3} \mathrm{B} -6709)
r^6+50648 \bar{s}^2 \nu_{3} r^2\\\nonumber &+11569 \bar{s}^2 \nu_{3}
^2\big)\big) d_3^2 r^4-4 \big(10240 \pi ^2 \bar{s}^2 \big(5 r^2+4
\nu_{3} \big) r^4+4 \pi \big( (11 \nu_{3} \mathrm{B} -10)
\\\nonumber &\times 256r^8+13360 \bar{s}^2 \nu_{3} r^4+2579
\bar{s}^2 \nu_{3} ^2 r^2\big)+\nu_{3} \big(1732 \bar{s}^2 \nu_{3}
r^2+222 \bar{s}^2 \nu_{3} ^2+r^6\\\nonumber &\times(2726 \nu_{3}
\mathrm{B} -1579) \big)\big) d_3^3 r^2+4 \big(2560 \pi ^2 \bar{s}^2
\big(20 r^2+3 \nu_{3} \big) r^4+4 \pi \big((3 \nu_{3} \mathrm{B}
-2)\\\nonumber &\times 240r^8+964 \bar{s}^2 \nu_{3} r^4+121
\bar{s}^2 \nu_{3} ^2 r^2\big)+\nu_{3} \big((26 \nu_{3} \mathrm{B}
+23) r^6+96 \bar{s}^2 \nu_{3} r^2\\\nonumber &+2 \bar{s}^2 \nu_{3}
^2\big)\big) d_3^4\big\} r^2+d_1^7 \big\{512 \pi ^2 \bar{s}^2
d_3^4-16 \pi \big((8 \nu_{3} \mathrm{B} -4) r^4+32 \pi \bar{s}^2
r^2+\bar{s}^2 \nu_{3}\\\nonumber &\times (31+16 \pi ) \big) d_3^3+2
\big(128 \pi ^2 r^2 \nu_{3} \bar{s}^2+4 \pi \big(56 \bar{s}^2
\nu_{3} r^2+31 \bar{s}^2 \nu_{3} ^2+(3 \nu_{3} \mathrm{B} -1)
\\\nonumber &\times 8r^6\big)+\nu_{3} \big(3 (8 \nu_{3} \mathrm{B}
-5) r^4+59 \bar{s}^2 \nu_{3} \big)\big) d_3^2-\nu_{3} \big(4 (18
\nu_{3} \mathrm{B} -5) r^6+2\bar{s}^2 \nu_{3} r^2\\\nonumber
&\times(51+112 \pi )+57 \bar{s}^2 \nu_{3} ^2\big) d_3+48 r^2
\bar{s}^2 \nu_{3} ^3\big\}+d_1^6 \big\{668 \bar{s}^2 \nu_{3} ^3
r^4-2 \nu_{3} \big(40r^6\\\nonumber &\times (11 \nu_{3} \mathrm{B}
-5)+2 (317+784 \pi ) \bar{s}^2 \nu_{3} r^2+421 \bar{s}^2 \nu_{3}
^2\big) d_3 r^2+64 \pi \big(104 \pi \bar{s}^2 r^2\\\nonumber
&-r^4+(1+4 \pi ) \bar{s}^2 \nu_{3} \big) d_3^4-2 \big(128 \pi ^2 r^2
\big(26 r^2+15 \nu_{3} \big) \bar{s}^2+4 \pi \big((19 \nu_{3}
\mathrm{B} -15)\\\nonumber &\times 8r^6+808 \bar{s}^2 \nu_{3} r^2+35
\bar{s}^2 \nu_{3} ^2\big)+\nu_{3} \big((66 \nu_{3} \mathrm{B} -43)
r^4+2 \bar{s}^2 \nu_{3} \big)\big) d_3^3+\big(3584\\\nonumber
&\times \pi ^2 \bar{s}^2 \nu_{3} r^4+8 \pi \big(8 (31 \nu_{3}
\mathrm{B} -14) r^8+728 \bar{s}^2 \nu_{3} r^4+463 \bar{s}^2 \nu_{3}
^2 r^2\big)+\nu_{3}\big(r^6\\\nonumber &\times (736 \nu_{3}
\mathrm{B} -602)+1460 \bar{s}^2 \nu_{3} r^2+69 \bar{s}^2 \nu_{3}
^2\big)\big) d_3^2\big\}+d_1^5 \big\{3982 \bar{s}^2 \nu_{3} ^3 r^6-2
\nu_{3}\\\nonumber &\times \big(4 (583 \nu_{3} \mathrm{B} -333)
r^6+(3483+9352 \pi ) \bar{s}^2 \nu_{3} r^2+2660 \bar{s}^2 \nu_{3}
^2\big) d_3 r^4+\big(21248\\\nonumber &\times \pi ^2 \bar{s}^2
\nu_{3} r^4+3 \nu_{3} \big(10 (148 \nu_{3} \mathrm{B} -131) r^6+2738
\bar{s}^2 \nu_{3} r^2+277 \bar{s}^2 \nu_{3} ^2\big)+8 \pi\\\nonumber
&\times \big(8 (150 \nu_{3} \mathrm{B} -83) r^8+4060 \bar{s}^2
\nu_{3} r^4+2925 \bar{s}^2 \nu_{3} ^2 r^2\big)\big) d_3^2 r^2+8
\big(32 \pi ^2 r^2\bar{s}^2\\\nonumber &\times \big(144 r^2+11
\nu_{3} \big) +\nu_{3} \big((5-12 \nu_{3} \mathrm{B} ) r^4+3
\bar{s}^2 \nu_{3} \big)+4 \pi \big((20 \nu_{3} \mathrm{B} -22)
r^6\\\nonumber &+31 \bar{s}^2 \nu_{3} r^2+\bar{s}^2 \nu_{3}
^2\big)\big) d_3^4-4 \big(128 \pi ^2 \bar{s}^2 \big(72 r^2+47
\nu_{3} \big) r^4+\nu_{3} \big( (64 \nu_{3} \mathrm{B} -41)
\\\nonumber &\times4r^6+84 \bar{s}^2 \nu_{3} r^2+3 \bar{s}^2 \nu_{3}
^2\big)+4 \pi \big(4 (99 \nu_{3} \mathrm{B} -94) r^8+2275 \bar{s}^2
\nu_{3} r^4\\\label{A3} &+204 \bar{s}^2 \nu_{3} ^2 r^2\big)\big)
d_3^3\big\}\big].
\end{align}

\end{document}